\documentclass[12pt, draftclsnofoot, onecolumn]{IEEEtran}

\usepackage{fancyhdr}
\usepackage{epsfig}
\usepackage{threeparttable}
\usepackage{epsf,epsfig}
\usepackage{amsthm}
\usepackage[cmex10]{amsmath}
\usepackage{amssymb, amsfonts}
\usepackage[noadjust]{cite}
\usepackage{dsfont}
\usepackage{subfigure}
\usepackage{color}
\usepackage{soul}
\usepackage{amssymb}
 \usepackage{accents}
  \usepackage{algorithm}
   \usepackage[english]{babel}
    \usepackage{url}
    \usepackage{framed}

\begin{document}
\newtheorem{theorem}{Theorem}
\newtheorem{acknowledgement}[theorem]{Acknowledgement}
\newtheorem{axiom}[theorem]{Axiom}
\newtheorem{case}[theorem]{Case}
\newtheorem{claim}[theorem]{Claim}
\newtheorem{conclusion}[theorem]{Conclusion}
\newtheorem{condition}[theorem]{Condition}
\newtheorem{conjecture}[theorem]{Conjecture}
\newtheorem{criterion}[theorem]{Criterion}
\newtheorem{definition}{Definition}
\newtheorem{exercise}[theorem]{Exercise}
\newtheorem{lemma}{Lemma}
\newtheorem{corollary}{Corollary}
\newtheorem{notation}[theorem]{Notation}
\newtheorem{problem}[theorem]{Problem}
\newtheorem{proposition}{Proposition}
\newtheorem{scheme}{Scheme}   
\newtheorem{policy}{Policy}   
\newtheorem{solution}[theorem]{Solution}
\newtheorem{summary}[theorem]{Summary}
\newtheorem{assumption}{Assumption}
\newtheorem{example}{\bf Example}
\newtheorem{remark}{\bf Remark}

\def\qed{$\Box$}
\def\QED{\mbox{\phantom{m}}\nolinebreak\hfill$\,\Box$}
\def\proof{\noindent{\emph{Proof:} }}
\def\poof{\noindent{\emph{Sketch of Proof:} }}
\def
\endproof{\hspace*{\fill}~\qed
\par
\endtrivlist\unskip}
\def\endproof{\hspace*{\fill}~\qed\par\endtrivlist\vskip3pt}

\def\E{\mathsf{E}}
\def\eps{\varepsilon}
\def\phi{\varphi}
\def\Lsp{{\boldsymbol L}}
\def\Bsp{{\boldsymbol B}}
\def\lsp{{\boldsymbol\ell}}
\def\Ltsp{{\Lsp^2}}
\def\Lpsp{{\Lsp^p}}
\def\Linsp{{\Lsp^{\infty}}}
\def\LtR{{\Lsp^2(\Rst)}}
\def\ltZ{{\lsp^2(\Zst)}}
\def\ltsp{{\lsp^2}}
\def\ltZt{{\lsp^2(\Zst^{2})}}
\def\ninN{{n{\in}\Nst}}
\def\oh{{\frac{1}{2}}}
\def\grass{{\cal G}}
\def\ord{{\cal O}}
\def\dist{{d_G}}
\def\conj#1{{\overline#1}}
\def\ntoinf{{n \rightarrow \infty }}
\def\toinf{{\rightarrow \infty }}
\def\tozero{{\rightarrow 0 }}
\def\trace{{\operatorname{trace}}}
\def\ord{{\cal O}}
\def\UU{{\cal U}}
\def\rank{{\operatorname{rank}}}
\def\acos{{\operatorname{acos}}}

\def\SINR{\mathsf{SINR}}
\def\SNR{\mathsf{SNR}}
\def\SIR{\mathsf{SIR}}
\def\tSIR{\widetilde{\mathsf{SIR}}}
\def\Ei{\mathsf{Ei}}
\def\l{\left}
\def\r{\right}
\def\({\left(}
\def\){\right)}
\def\lb{\left\{}
\def\rb{\right\}}

\setcounter{page}{1}

\newcommand{\eref}[1]{(\ref{#1})}
\newcommand{\fig}[1]{Fig.\ \ref{#1}}

\def\bydef{:=}
\def\ba{{\mathbf{a}}}
\def\bb{{\mathbf{b}}}
\def\bc{{\mathbf{c}}}
\def\bd{{\mathbf{d}}}
\def\bee{{\mathbf{e}}}
\def\bff{{\mathbf{f}}}
\def\bg{{\mathbf{g}}}
\def\bh{{\mathbf{h}}}
\def\bi{{\mathbf{i}}}
\def\bj{{\mathbf{j}}}
\def\bk{{\mathbf{k}}}
\def\bl{{\mathbf{l}}}
\def\bm{{\mathbf{m}}}
\def\bn{{\mathbf{n}}}
\def\bo{{\mathbf{o}}}
\def\bp{{\mathbf{p}}}
\def\bq{{\mathbf{q}}}
\def\br{{\mathbf{r}}}
\def\bs{{\mathbf{s}}}
\def\bt{{\mathbf{t}}}
\def\bu{{\mathbf{u}}}
\def\bv{{\mathbf{v}}}
\def\bw{{\mathbf{w}}}
\def\bx{{\mathbf{x}}}
\def\by{{\mathbf{y}}}
\def\bz{{\mathbf{z}}}
\def\b0{{\mathbf{0}}}

\def\bA{{\mathbf{A}}}
\def\bB{{\mathbf{B}}}
\def\bC{{\mathbf{C}}}
\def\bD{{\mathbf{D}}}
\def\bE{{\mathbf{E}}}
\def\bF{{\mathbf{F}}}
\def\bG{{\mathbf{G}}}
\def\bH{{\mathbf{H}}}
\def\bI{{\mathbf{I}}}
\def\bJ{{\mathbf{J}}}
\def\bK{{\mathbf{K}}}
\def\bL{{\mathbf{L}}}
\def\bM{{\mathbf{M}}}
\def\bN{{\mathbf{N}}}
\def\bO{{\mathbf{O}}}
\def\bP{{\mathbf{P}}}
\def\bQ{{\mathbf{Q}}}
\def\bR{{\mathbf{R}}}
\def\bS{{\mathbf{S}}}
\def\bT{{\mathbf{T}}}
\def\bU{{\mathbf{U}}}
\def\bV{{\mathbf{V}}}
\def\bW{{\mathbf{W}}}
\def\bX{{\mathbf{X}}}
\def\bY{{\mathbf{Y}}}
\def\bZ{{\mathbf{Z}}}

\def\mA{{\mathbb{A}}}
\def\mB{{\mathbb{B}}}
\def\mC{{\mathbb{C}}}
\def\mD{{\mathbb{D}}}
\def\mE{{\mathbb{E}}}
\def\mF{{\mathbb{F}}}
\def\mG{{\mathbb{G}}}
\def\mH{{\mathbb{H}}}
\def\mI{{\mathbb{I}}}
\def\mJ{{\mathbb{J}}}
\def\mK{{\mathbb{K}}}
\def\mL{{\mathbb{L}}}
\def\mM{{\mathbb{M}}}
\def\mN{{\mathbb{N}}}
\def\mO{{\mathbb{O}}}
\def\mP{{\mathbb{P}}}
\def\mQ{{\mathbb{Q}}}
\def\mR{{\mathbb{R}}}
\def\mS{{\mathbb{S}}}
\def\mT{{\mathbb{T}}}
\def\mU{{\mathbb{U}}}
\def\mV{{\mathbb{V}}}
\def\mW{{\mathbb{W}}}
\def\mX{{\mathbb{X}}}
\def\mY{{\mathbb{Y}}}
\def\mZ{{\mathbb{Z}}}

\def\cA{\mathcal{A}}
\def\cB{\mathcal{B}}
\def\cC{\mathcal{C}}
\def\cD{\mathcal{D}}
\def\cE{\mathcal{E}}
\def\cF{\mathcal{F}}
\def\cG{\mathcal{G}}
\def\cH{\mathcal{H}}
\def\cI{\mathcal{I}}
\def\cJ{\mathcal{J}}
\def\cK{\mathcal{K}}
\def\cL{\mathcal{L}}
\def\cM{\mathcal{M}}
\def\cN{\mathcal{N}}
\def\cO{\mathcal{O}}
\def\cP{\mathcal{P}}
\def\cQ{\mathcal{Q}}
\def\cR{\mathcal{R}}
\def\cS{\mathcal{S}}
\def\cT{\mathcal{T}}
\def\cU{\mathcal{U}}
\def\cV{\mathcal{V}}
\def\cW{\mathcal{W}}
\def\cX{\mathcal{X}}
\def\cY{\mathcal{Y}}
\def\cZ{\mathcal{Z}}
\def\cd{\mathcal{d}}
\def\Mt{M_{t}}
\def\Mr{M_{r}}
\def\O{\Omega_{M_{t}}}
\newcommand{\figref}[1]{{Fig.}~\ref{#1}}
\newcommand{\tabref}[1]{{Table}~\ref{#1}}

\newcommand{\var}{\mathsf{var}}
\newcommand{\fb}{\tx{fb}}
\newcommand{\nf}{\tx{nf}}
\newcommand{\BC}{\tx{(bc)}}
\newcommand{\MAC}{\tx{(mac)}}
\newcommand{\Pout}{p_{\mathsf{out}}}
\newcommand{\nnn}{\nn\\}
\newcommand{\FB}{\tx{FB}}
\newcommand{\TX}{\tx{TX}}
\newcommand{\RX}{\tx{RX}}
\renewcommand{\mod}{\tx{mod}}
\newcommand{\m}[1]{\mathbf{#1}}
\newcommand{\td}[1]{\tilde{#1}}
\newcommand{\sbf}[1]{\scriptsize{\textbf{#1}}}
\newcommand{\stxt}[1]{\scriptsize{\textrm{#1}}}
\newcommand{\suml}[2]{\sum\limits_{#1}^{#2}}
\newcommand{\sumlk}{\sum\limits_{k=0}^{K-1}}
\newcommand{\eqhsp}{\hspace{10 pt}}
\newcommand{\tx}[1]{\texttt{#1}}
\newcommand{\Hz}{\ \tx{Hz}}
\newcommand{\sinc}{\tx{sinc}}
\newcommand{\tr}{\mathrm{tr}}
\newcommand{\diag}{\mathrm{diag}}
\newcommand{\MAI}{\tx{MAI}}
\newcommand{\ISI}{\tx{ISI}}
\newcommand{\IBI}{\tx{IBI}}
\newcommand{\CN}{\tx{CN}}
\newcommand{\CP}{\tx{CP}}
\newcommand{\ZP}{\tx{ZP}}
\newcommand{\ZF}{\tx{ZF}}
\newcommand{\SP}{\tx{SP}}
\newcommand{\MMSE}{\tx{MMSE}}
\newcommand{\MINF}{\tx{MINF}}
\newcommand{\RC}{\tx{MP}}
\newcommand{\MBER}{\tx{MBER}}
\newcommand{\MSNR}{\tx{MSNR}}
\newcommand{\MCAP}{\tx{MCAP}}
\newcommand{\vol}{\tx{vol}}
\newcommand{\ah}{\hat{g}}
\newcommand{\tg}{\tilde{g}}
\newcommand{\teta}{\tilde{\eta}}
\newcommand{\heta}{\hat{\eta}}
\newcommand{\uh}{\m{\hat{s}}}
\newcommand{\eh}{\m{\hat{\eta}}}
\newcommand{\hv}{\m{h}}
\newcommand{\hh}{\m{\hat{h}}}
\newcommand{\Po}{P_{\mathrm{out}}}
\newcommand{\Poh}{\hat{P}_{\mathrm{out}}}
\newcommand{\Ph}{\hat{\gamma}}
\newcommand{\mat}[1]{\begin{matrix}#1\end{matrix}}
\newcommand{\ud}{^{\dagger}}
\newcommand{\C}{\mathcal{C}}
\newcommand{\nn}{\nonumber}
\newcommand{\nInf}{U\rightarrow \infty}

\title{\huge  Data-Importance Aware User Scheduling for Communication-Efficient Edge Machine Learning } 

\author{Dongzhu Liu, Guangxu Zhu, Jun Zhang, and Kaibin Huang
\thanks{\noindent  D. Liu, G. Zhu, and K. Huang are with the Dept. of Electrical and Electronic Engineering at The University of Hong Kong, Hong Kong. J. Zhang is with the Dept. of Electronic and Information Engineering at the Hong Kong Polytechnic University, Hong Kong. Corresponding author: K. Huang (email: huangkb@eee.hku.hk). 
}}

\maketitle

\vspace{-50pt} 

\begin{abstract} 
With the prevalence of intelligent mobile applications,  edge learning is emerging as a  promising technology for powering fast intelligence acquisition for edge devices from distributed data generated at the network edge.  
One critical task of edge learning is to efficiently utilize the limited radio resource to acquire data samples for model training at an edge server. In this paper, we develop a novel user scheduling algorithm for data acquisition in edge learning, called \emph{(data) importance-aware scheduling}.  A key feature of this scheduling algorithm is that it takes into account the informativeness of data samples, besides communication reliability.  Specifically, the scheduling decision is  based on a \emph{data importance indicator} (DII), elegantly incorporating two ``important" metrics from communication and learning perspectives, i.e., the \emph{signal-to-noise ratio} (SNR) and \emph{data uncertainty}.  We first derive an explicit expression for this indicator targeting the classic classifier of \emph{support vector machine} (SVM), where the uncertainty of a data sample is measured by its distance to the decision boundary. Then, the result is extended to \emph{convolutional neural networks} (CNN) by replacing the distance based uncertainty measure with the entropy.  As demonstrated via experiments using real datasets, the proposed importance-aware scheduling can exploit the two-fold multi-user diversity, namely the diversity in both the multiuser channels and the distributed data samples.  This leads to faster model convergence than the conventional scheduling schemes that exploit only a single type of diversity.  
\end{abstract} 

\vspace{-15pt}
\section{Introduction}
The proliferation of smart devices and the booming of \emph{artificial intelligence}  (AI) ushered in a new era of ambient intelligence. Materializing the vision motivates the deployment of machine learning algorithms at the network edge, named \emph{edge learning} \cite{zhu2018towards, wang2018edge, park2018wireless, zhou2019edge, zhang2019mobile}, to enable intelligent mobile applications.  Edge learning aims at fast AI model training by exploiting computing resources at edge servers, and low-latency access to distributed data at edge devices.  In return, downloading the trained models to the devices will equip them with human-like intelligence to cope with real-time inference and decision making.  Edge learning sits at the intersection of two areas: wireless communications and machine learning.  The emergence of the new area gives rise to many interdisciplinary research opportunities that require joint designs interweaving the two said areas towards an ultimate goal of fast and efficient intelligence acquisition.

With rapidly growing data-processing speeds, the bottleneck of fast edge learning is more on the communication aspect.  Specifically, wirelessly uploading  high-dimensional data from a large number of edge devices can congest the air-interface due to the limited radio resource  \cite{bonawitz2019towards}.  
To overcome this bottleneck,  it calls for innovations on highly efficient wireless data acquisition tailored for edge learning systems.  The conventional wireless technologies focus on \emph{rate maximization} or \emph{Quality-of-Service (QoS)}, which implicitly assume that transmitted data bits are equally important.  However, the assumption is improper for machine learning applications since some data samples are more effective than the others for improving a learning model \cite{settles2012active}. 
This fact motivates a novel design principle of importance aware \emph{radio resource management} (RRM)  that the radio resource should be allocated to edge devices not only based on channel states but also accounting for the importance of  their data for model training. In this work, we apply this principle to revamp user scheduling by exploiting multi-user diversity in both the channel and data domains for efficient wireless data acquisition.

\subsection{Related Work and Motivation}  

 \subsubsection{Radio Resource Management for E Edge Learning}
 
 The mission of conventional wireless com- munications is to reliably transmit data bits at a rate as high as possible, regardless of the data content and its usefulness. Therefore, directly applying such a communication-learning separation principle to edge learning will lead to inefficient transmission \cite{popovski2019semantic}. This has triggered a lot of research interests recently on redesigning communication techniques for edge learning [1], covering key topics such as RRM \cite{konevcny2016federated,mcmahan2016comm, ren2019accelerating,skatchkovsky2019optimizing, chen2018label,qian2019active, liu2018wireless}, multiple access \cite{zhu2018low,amiri2019machine,yang2018federated}, and signal encoding  \cite{alistarh2017qsgd, horvath2019stochastic, du2018fast}. The new idea in RRM, the theme of this paper, is to allocate resources to edge devices for transmitting learning relevant data by considering the learning task. Prior work on this topic can be separated for  two learning paradigms. The first paradigm is federated edge learning that preserves privacy by avoiding direct data uploading. In this paradigm, a model is distributively trained at edge devices, and the purpose of uplink transmission is to upload local models which are aggregated at an edge server to cooperatively improve a global model \cite{konevcny2016federated,mcmahan2016comm}. In this paradigm, a resource allocation method integrating computing and communication is proposed to improve the learning efficiency \cite{ren2019accelerating}. Specifically, the training batch size is adapted to the wireless channel condition for attaining higher learning accuracy without compromising the latency. A similar idea has also been investigated in the other paradigm, centralized edge learning, where edge devices directly upload data to the server for training the global model. In this paradigm, given the communication overhead, the offloading data size in each communication round is optimized to acquire sufficient data samples for reducing the learning bias, and avoids insufficient learning due to exceeding the computing capacity, thereby improving the learning performance \cite{skatchkovsky2019optimizing}. The efficiency of wireless data acquisition can be further enhanced by differentiating the usefulness of training data samples \cite{chen2018label,qian2019active}. Specifically, a novel retransmission scheme is designed for adapting the reliability requirement of a received data sample according to its importance for model improvement \cite{liu2018wireless}. By intelligently allocating the constrained transmission budget, the scheme allows more important data to be received with a guaranteed reliability compared to the conventional channel-aware design, thereby improving learning accuracy. This idea of importance-aware retransmission motivates us to propose the new principle of importance-aware user scheduling to explore the new dimension of multiuser diversity in both channels and data to improve the communication efficiency of edge learning.

 \subsubsection{Multiuser Diversity}
Multiuser channel diversity is an intrinsic characteristic of wireless networks arising from independent  fading in multiuser channels. To increase the network throughput, multiuser diversity can be exploited by scheduling the user with the \emph{best channel} at any given time \cite{knopp1995information,tse1998multiaccess}.
The diversity gain tends to increase for channels with large dynamic ranges, e.g.,  rich scattering and fast fading,
as well as the large number of users.
On the other hand, the scheme targeting throughput maximization is biased towards users with favourable channels and unfair for others \cite{liu2001opportunistic}.  To address this issue, one solution is proportional fair scheduling where a scheduling metric being the radio between  the instantaneous and average rates of each user is adopted to strike a balance between rate maximization and fairness \cite{liu2003opportunistic, kwan2009proportional,bang2008channel}.  In the existing work,  data importance is assumed homogeneous.  However, in the context of edge learning, data samples differ in their importance for learning, called \emph{data diversity}.  Then the distribution of data at multiple devices gives rise to a new type of multiuser diversity, namely, \emph{multiuser data diversity}.  In this work, we make the first attempt to exploit both types of multiuser diversity in scheduling so as to improve the communication efficiency of edge learning.


Data diversity is not new but a fundamental concept in the area of \emph{active learning}~\cite{settles2012active}.  It concerns a scenario where abundant unlabelled data are available and manual labelling is costly.   Data diversity can be exploited by selecting the \emph{most informative data samples} to be labeled  (by querying an oracle), such that a model can be accurately trained using fewer labelled samples, thereby reducing the labelling cost.
Generally, the informative data samples are those highly uncertain to be predicted under the current model.  Their use in training can significantly improve the accuracy of the classifier model. 
There are several commonly adopted metrics for measuring data uncertainty including  \emph{entropy} \cite{holub2008entropy}, \emph{expected model change}~\cite{settles2008multiple}, and \emph{expected error reduction}~\cite{roy2001toward}.  
For active learning, all data are assumed to be located at a server and hence wireless transmission is irrelevant.  Nevertheless, the data-uncertainty measures developed in the area are found in this work to be a useful tool for designing importance aware scheduling for wireless data acquisition in edge learning.

\subsection{Contributions and Organization} 
In this work, we propose importance-aware scheduling for communication efficient edge learning. To this end, consider a centralized edge learning system where a classifier is trained at the edge server by utilizing the data distributed at multiple edge devices. To accelerate the model training,  the edge server schedules a device for wireless data uploading under the criterion of maximum improvement on the classifier's accuracy.  
The proposed scheduling scheme exploits channel diversity and data diversity simultaneously, to ensure received data are both important and reliable in the presence of channel fading and noise. As a result, the model convergence is accelerated and channel uses reduced. To the authors' best knowledge, this work represents the first attempt on exploiting both the channel and data diversity to improve the communication efficiency of an edge learning system.

The main contributions of this work are summarized as follows. 

\begin{itemize}
\item{\bf Importance-aware scheduling for SVM:} We first consider the classic classifier  model of \emph{support vector machine} (SVM), and develop the basic principle of importance-aware scheduling. The core element of the scheme is a new scheduling metric, named \emph{data importance indicator} (DII), that is proposed to be the expected uncertainty of a received data sample in the presence of channel fading and noise.  For SVM, the DII is suitably defined as the expected negative distance from a received data sample to the decision boundary of the classifier.  The theoretical contribution of the DII design lies in that its derived closed form elegantly combines the received \emph{signal-to-noise ratio} (SNR) from the communication perspective and  \emph{data uncertainty}  from the learning perspective in a simple addition form.  This allows the DII to measure the effective importance of a received data sample for model training given fading and noise.  Consequently, scheduling under the criterion of maximum DII, yielding the proposed importance-aware scheduling,  is capable of exploiting both multiuser channel-and-data diversity to accelerate model convergence while effectively coping with channel hostility.  

\item{\bf  Extension to general classifiers:} The principle of importance-aware scheduling developed for SVM is extended to general classifier models.  
The generalization essentially replaces the distance-based uncertainty measure in the previous DII design for SVM with a general measure.  It can be specified as one of available measures (e.g., entropy or expected model change) depending on the design choice.   
For illustration, a case study  for the modern \emph{convolutional neural networks} (CNN) classifier is presented.  

\item{\bf  Practical issues in implementation:}  Several practical implementation issues of the proposed scheme are discussed and addressed by suitable design modifications.  
\begin{itemize}
\item{\bf Exploiting data-label information:} The design involves revising the DII as the expectation of model update, which is derived to combine the SNR and label dependent hinge loss (see e.g., \cite{bishop2006pattern}) in a product form.  Based on the revised DII with label information, the scheduling scheme can achieve faster convergence rate than the previous design.  
\item{\bf Model compression:} The second issue is the high computational complexity of data uncertainty evaluation at edge devices.  This can be addressed by using a compressed model that prunes the model parameters with small values. 
\item{\bf Data deficiency:} In practice, the scheduling may face the data deficiency due to limited available devices in the system.  That may degrade the performance of the proposed scheme since it is highly dependent on the global data size  (or higher data diversity).  To cope with this issue, several practical solutions are discussed for increasing the data size by increasing the number of devices, increasing buffer sizes, updating the local buffer with a higher frequency, or utilizing user mobility.  
\end{itemize}

\item{\bf  Experiments:}  We evaluate the performance of the proposed importance-aware scheduling via extensive experiments using real datasets. 
The results demonstrate that the proposed method is able to exploit the two types of multiuser diversity, and as a result achieves better learning performance than the two baseline schemes that exploit only a single type of diversity. Moreover, the performance can be further improved by increasing the data size using several proposed  methods.
By exploiting the label information, the importance-aware scheduling could attain a faster convergence rate.  
Last, the computational complexity can be reduced by implementing a compressed evaluation model without significantly compromising the learning performance.

\end{itemize}

The remainder of the paper is organized as follows. The communication and learning models are introduced in Section~\ref{sec: system model}. In Section~\ref{sec: principle without label}, the principle of importance-aware scheduling is proposed.  Several practical issues and solutions are discussed in Section~\ref{sec: discussions}.  Section~\ref{sec:simulation} provides experimental results, followed by concluding remarks in Section~\ref{sec: concluding remarks}.

\section{ Communication and Learning Models}\label{sec: system model}
In this section, we first introduce the communication model, including multiuser scheduling and the data channel model. Then the learning models are introduced, followed by the data importance measures.  

\vspace{-10pt}
\subsection{Communication Model}
\begin{figure}[t]
\centering
\subfigure[Scheduling and data uploading.]{
\label{Fig: scheduling}
\includegraphics[width=13cm]{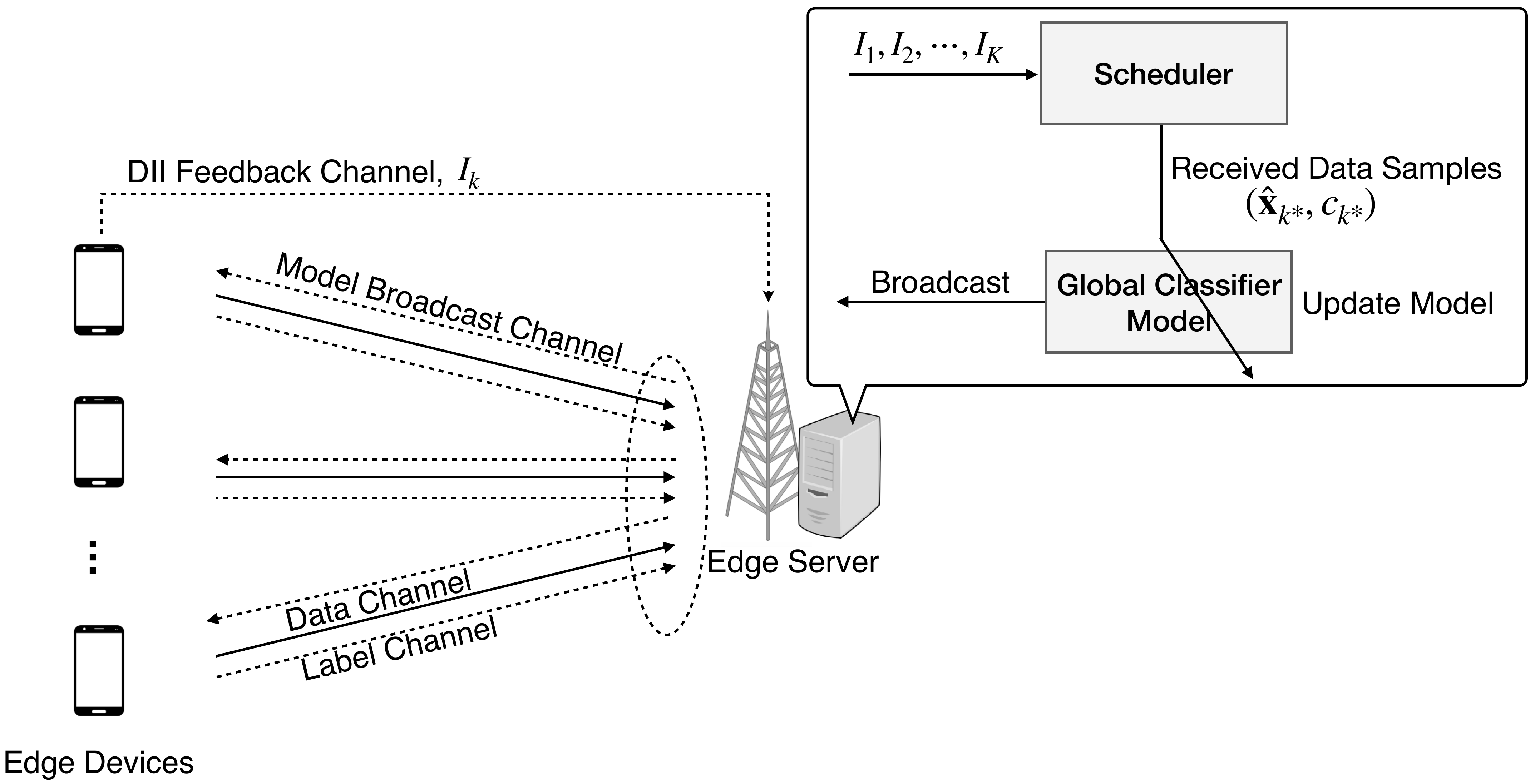}}
\subfigure[Data importance evaluation at an edge device.]{
\label{Fig: ImpEval}
\includegraphics[width=11cm]{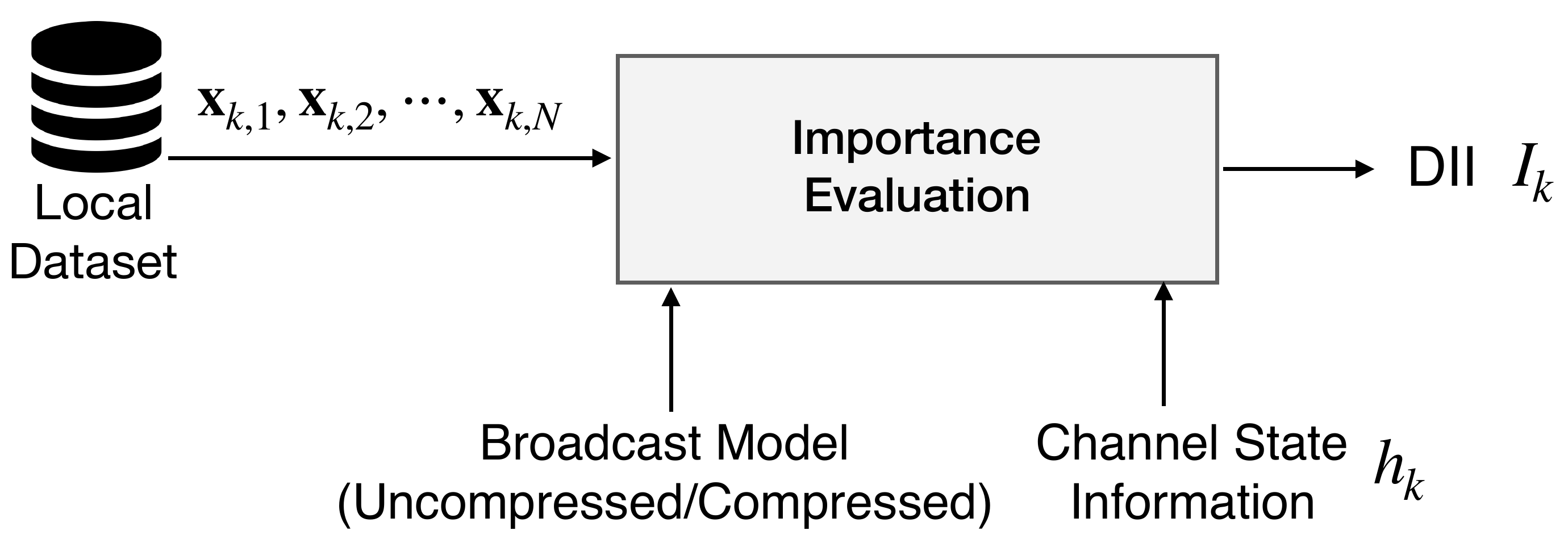}}
\caption{An edge learning system with importance-aware scheduling.}
\vspace{-20pt} 
\label{Fig: system}
\end{figure}

We consider an edge learning system in a single-cell wireless network as  shown in Fig.~\ref{Fig: system}, which comprises a single edge server and multiple edge devices, each equipped with a single antenna.  A machine learning model is to be trained at the edge sever by utilizing the labeled data samples distributed over the $K$ edge devices.  The devices are coordinated by a scheduler to share the wireless channel in a time division manner, and they take turn to upload a data sample in each time slot.   
Each device is equipped a local buffer with the size of $N$ samples.  
Note that both the buffer updating frequency and device mobility affect the learning performance which are discussed in Section~\ref{sec: mobility and update}.  
Denote the $n$-th data sample at the $k$-th device as $\bx_{k,n} \in \mathbb{R}^p$, {\color{black} and its label $c_k \in \{1, 2, \cdots, C\}$ is acquired after the data sample is selected for transmission.}
Note that a label has a much smaller size than a data sample (e.g., a $0-9$ integer versus a vector of a million real coefficients), thus a low-rate noiseless channel for label transmission is assumed for simplicity.


\subsubsection{Multiuser Scheduling} 
Time is divided into symbol durations, called \emph{slots}. Slot synchronization among devices are assumed.  
Transmission of a data sample requires $p$ slots, called a \emph{symbol block}.  At the beginning of each symbol block, the edge server broadcasts the global model for the devices to evaluate the importance of their data samples, measured by the DII and denoted as $I_k$ at the $k$-th device, based on which, a device is selected for data uploading. The main purpose of this work is to design the DII.  
The model broadcast for data importance evaluation can be the current global model under training or a compressed one for low-complexity computation, as discussed in Section~\ref{subsec: compressed model}. 
Assuming a noiseless broadcast model and perfect \emph{channel state information} (CSI), the data importance measure is evaluated at the devices and the results fed back to the server for scheduling.  
Upon receiving the DIIs, the server selects one of the devices for data-sample transmission.  
 
\subsubsection{Data Channel Model} Data channels are assumed to follow block-fading, where the channel coefficients remain static within a symbol block and are \emph{independent and identically distributed} (i.i.d.) over different users. In a series of recent studies, analog modulation is seeing its revival to be a promising solution for multimedia transmission and found to outperform its digital counterpart in terms of edge learning performance \cite{ImageTXGunduz2019}, in the presence of Gaussian noise \cite{KatabiAnalogTransmission2010}, in compression efficiency  \cite{LaiAnalogVideoBroadcast2014} and power consumption  \cite{NuPowerEfficientVideo2018} for video transmission, and in alleviating the noise effect on video quality \cite{KatabiSoftCast2010}. For these advantages as well as simplification of the design, 
analog modulation is adopted for transmitting training data which are  typically images or videos. Specifically, during an arbitrary symbol block, the scheduled $k$-th device sends the data sample ${ \bf x}$ using linear analog modulation, yielding the received signal given by
\begin{equation}\label{channel model}
{\bf y}=\sqrt{P}h_k{\bf x}+{\bf z}_k,
\end{equation} 
where $P$ is the transmit  power, the Rayleigh fading coefficient $h_k$ is a complex Gaussian random variable (r.v.), i.e., $h_k\sim{\cal CN}(0,1)$,  and ${\bf z}_k$ is the \emph{additive white Gaussian noise} (AWGN) vector with the entries  following the i.i.d. ${\cal CN}(0,\sigma^2)$ distributions. Analog uncoded transmission is assumed here to allow fast data transmission \cite{amiri2019machine} and for a higher energy efficiency (compared with the digital counterpart) as pointed out by \cite{cui2005energy}.   We assume that perfect CSI is available at the edge server. This allows the server to compute  the instantaneous SNR and decode the received sample $\hat{{ \bf x}}$ as follows:
\begin{equation} 
\hat{{ \bf x}}=\frac{1}{\sqrt{P}}\Re\l(\frac{{h_k}^*{\bf y}}{\|{h_k}\|^2}\r),\label{eq: est x non comb}
\end{equation}
where ${\bf y}$ is given in \eqref{channel model}. In \eqref{eq: est x non comb}, we extract the real part of the combined signal for further processing since the data for machine learning  are real-valued in general (e.g., photos, voice clips or video clips).  As a result, the  receive SNR for sample $\hat{{ \bf x}}$ is given as 
\begin{equation}\label{eq: def SNR non comb}
\SNR_k=\frac{2P}{\sigma^2}|h_k|^2,
\end{equation}
where the coefficient $2$ at  the right hand side arises from the fact that only the noise in the real dimension with variance  $\frac{\sigma^2}{2}$  affects the received data. The SNR expression in \eqref{eq: def SNR non comb} measures the reliability of a received data sample and serves as one of two performance metrics to be accounted for in scheduling as discussed in Section~\ref{sec: principle without label}.


\subsection{Learning Model}

For the learning task, we consider supervised training of a classifier.  Prior to wireless data acquisition, there is some initial data at the server.
The data, denoted as ${\cal L}_0$, allow the construction of a coarse initial classifier, which is used for importance evaluation at the beginning.  The classifier is refined  progressively in the subsequent data acquisition (and training) process.   In this work, we consider two widely used classifier models, i.e., the classic SVM classifier and the modern CNN classifier as introduced below.

\begin{figure}[t]
\begin{center}
{\includegraphics[width=12cm]{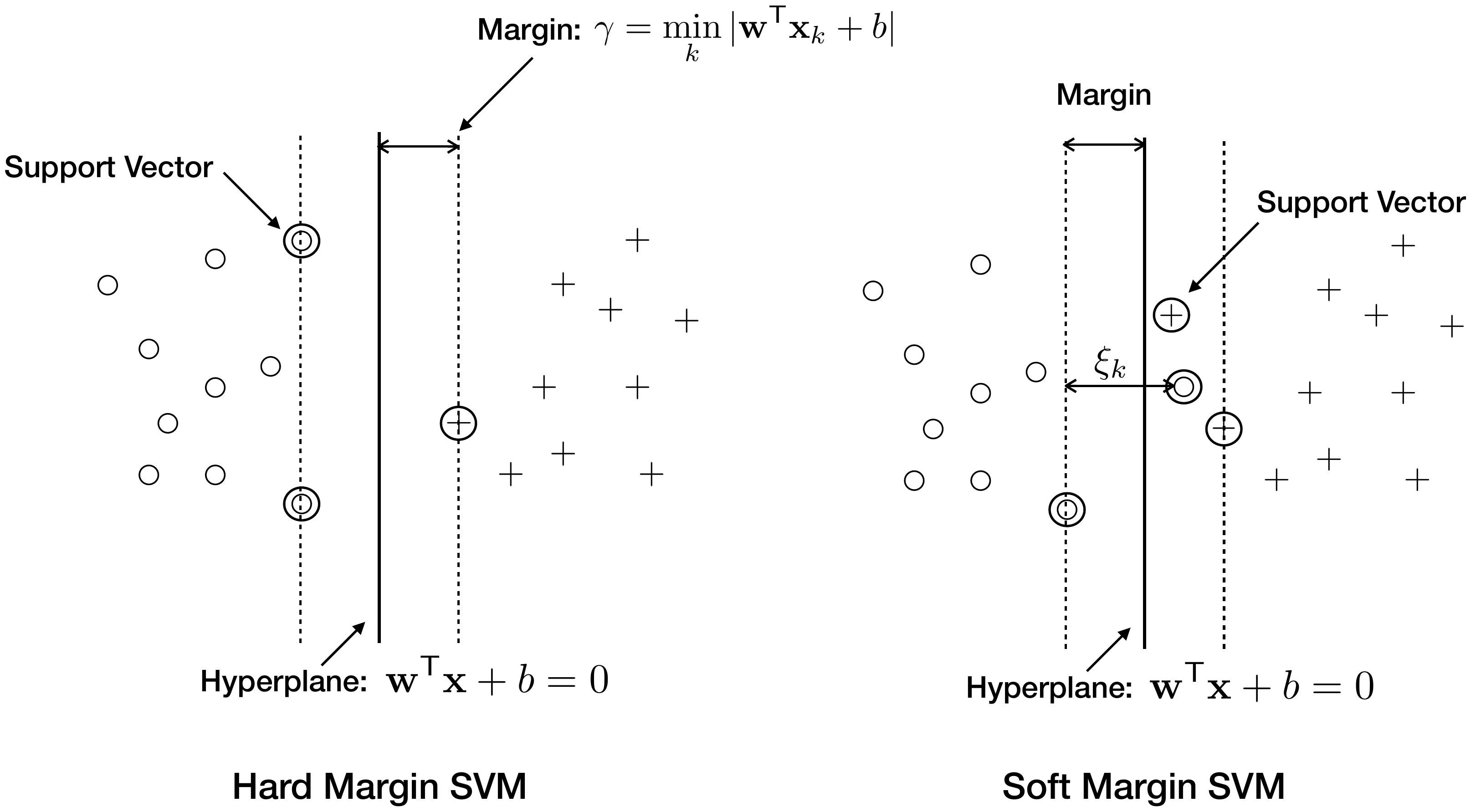}}\\
\vspace{-10pt} 
\caption{Comparison of hard-margin SVM and soft-margin SVM for binary classification.}
\vspace{-30pt} 
\label{Fig: SVM}
\end{center}
\end{figure}

\subsubsection{SVM Model}  As shown in Fig.~\ref{Fig: SVM}, the original \emph{hard margin SVM} is to seek the optimal hyperplane ${\bf w} ^{\sf T}{\bf x}+b=0$ as a decision boundary by maximizing its margin $\gamma$ to data points, i.e., the minimum distance between the hyperplane to any data sample \cite{friedman2001elements}. However, it works only for linearly separable datasets, which is hardly the case when the dataset is corrupted by channel noise in the current scenario. To enable the algorithm to cope with a potential outlier caused by noise, a variant of SVM called \emph{soft-margin SVM} is adopted. Soft-margin SVM  is widely used in practice to classify a noisy dataset that is not linearly separable by allowing misclassification, but with an additional penalty $\xi_i$ for the non-separable sample $\bx_i$. The comparison between {hard margin SVM} and {soft margin SVM} is graphically shown in Fig.~\ref{Fig: SVM}. A convex formulation for the soft margin SVM problem is given by
\begin{align}
\min_{{\bf w},b}\;\;  & \frac{1}{2}\|{\bf w} \|^2 + C \sum_{i} \xi_i\nn\\
{\rm s.t.} \;\;&c_i({\bf w} ^{\sf T}{\bf x}_i+b)\geq 1-\xi_i,   \label{problem: soft SVM}\\
& \xi_i\geq 0,  \quad   \forall i, \nn
\end{align} 
where $C$ is a parameter to control the tradeoff between maximizing the margin and minimizing the training error.  Small $C$ tends to emphasize the margin maximization by allowing certain level of misclassification in the training data, and vice versa. 
In soft-margin SVM, as shown in~Fig.~\ref{Fig: SVM}, the support vector is defined to be the point lies either on or inside the margin  \cite{bishop2006pattern}.  
Mathematically, a labelled training data sample $(\bx,c)$ is a support vector if it satisfies the following equation:   
\begin{equation}\label{eq: sv}
 \text{({\bf Support Vector}) }\ V(\bx,c)=1-c({\bf w} ^{\sf T}{\bf x}+b)\geq0.
\end{equation}
After training, the learnt SVM model can be used for predicting the label of a new sample by computing its output score is given as 
\begin{equation}\label{eq: redef score}
\text{({\bf Output  Score}) }\quad s({\bf x})=({\bf w} ^{\sf T}{\bf x}+b)/\|\bw\|,
\end{equation}
where $\| \cdot\|$ represents the Euclidean norm.  


%


\subsubsection{CNN model} CNN is made up of neurons that have adjustable weights and biases to express a non-linear mapping from an input data sample to class scores as outputs \cite{haykin1994neural}.   Note that the weights and biases constitute the parameters of the  CNN.  Typical implementation of a CNN consists of multiple layers including   convolutional layers, ReLu layers, pooling layers, fully connected layers and normalization layers. 
Without the explicitly defined decision boundaries as for SVM, CNN adjusts the parameters of the hidden layers to minimize the prediction error, calculated using the outputs of the softmax layer and the true labels of training data.  The expression of output score is given as: 
\begin{equation}\label{eq: redef score CNN}
\text{({\bf Output  Score}) }\quad s_{\hat c}({\bf x})=P_{\theta}(\hat c|\bx),
\end{equation}
indicating the posterior distribution of the predicted label of a data sample.  
After training, the learnt CNN model can then be used for predicting the label of a new sample by choosing one with the highest posterior probability.


\subsection{Data Uncertainty Measures}

The importance of a data sample for learning is usually measured by its \emph{uncertainty}, as viewed by the model under training \cite{settles2012active}.  Two uncertainty measures targeting SVM and CNN respectively are introduced as follows.

\subsubsection{Uncertainty Measure for SVM} For SVM, we adopt the distanced based uncertainty which  is motivated by the fact that a classifier makes less confident inference on a  data sample which is located  near the decision boundary \cite{liu2018wireless}.  Given a data sample ${\bf x}$ and a binary classifier $\{\bw , b\}$, the said distance can be readily computed by the absolute value of the output score as follows 
\begin{equation}
 d({\bf x}) =  | s({\bf x})|  = |{\bf w} ^{\sf T}{\bf x}+b|/ \|\bw\|. \label{eq:dis}
\end{equation}
Then the distance based uncertainty measure is defined as
\begin{equation}
\text{({\bf Distance Based Uncertainty}) }\quad   \mathcal{U}_{\sf d}\l({\bf x}\r) = -{ d^2({\bf x})} = -\frac{({\bf w} ^{\sf T}{\bf x}+b)^2}{\|\bw\|^2}. \label{eq: dis un}
\end{equation}

\subsubsection{Uncertainty Measure for CNN} For CNN, a suitable measure is \emph{entropy}, an information theoretic notion, defined as follows~\cite{holub2008entropy}:
\begin{equation}
\text{({\bf Entropy}) }\quad  \mathcal{U}_{\sf e}\l({\bf x}\r)=-\sum_{\hat c=1}^{C}P_{\theta}\l( \hat c | \mathbf{ x}\r)\log P_{\theta}\l( \hat c | \mathbf{ x}\r), \label{eq: entropy un}
\end{equation}
where $\hat{c}$ denotes a predicted class label and $\theta$ the set of model parameters to be learnt. 

\section{Principle of Importance-Aware Scheduling}\label{sec: principle without label}

{\color{black}
In this section, we first consider the task of training a binary SVM classifier at the edge. To attain a more accurate model under the constrained transmission budget, it requires the edge server to schedule the device with the most useful data sample for transmission.  The scheduling decision making is challenging as there lacks a selection metric to evaluate the importance of noisy data.  The problem is tackled in this section by designing the DII, which combines two metrics in communication and learning to indicate the effective importance of a transmitted data sample for learning.  Then, the importance-aware scheduling is proposed based on the indicator, so as to accelerate the model training at the edge server.  Finally, the proposed scheme for SVM is extended to general classifiers.

\vspace{-10pt}
\subsection{Data Importance Indicator}
The direct design of DII for optimizing the learning performance is difficult due to a lack of tractable mapping from noisy data importance to learning learning speed and accuracy. Nevertheless, the following fact in active learning provides a potential connection between data uncertainty and model-convergence rate: a model can be trained using  fewer labelled data samples if the highly uncertain data is selectively added into the training set.  The fact suggests that data uncertainty should be incorporated  into the design of DII.
However, an uncertainty measure in active learning targets noiseless data selection, and thus cannot be directly used for edge learning, as the acquired training data is corrupted by channel fading and noise, thereby affecting the effective uncertainty. 
To address this issue, the expectation of received data uncertainty can serve as a reasonable measure of effective uncertainty, and thus is used for defining the DII.  The definition is mathematically given as follows.
}



\begin{definition}[Data Importance Indicator]
\emph{Conditioned on the local dataset $D_k$ at the $k$-th edge device and its associated channel, the corresponding DII is defined as:
\begin{align}
I_k=\max_{n\in\mathcal{N}}\E_{\bz_k}\l[ \mathcal{U}_{\sf d}\l(\hat{\bf x}_{k,n}\r)\r],
\end{align} 
where $\hat{\bf x}_{k,n}$ and $ \mathcal{U}_{\sf d}(\cdot)$ are defined in \eqref{eq: est x non comb} and \eqref{eq: dis un} respectively, and $\mathcal{N}=\{1,2,\cdots,N\}$ represents the sample index set.}
\end{definition}

The remainder of the sub-section focuses on deriving a closed-from expression for  DII.  To begin with, we first give the expression for calculating the distance from a received data sample to the decision boundary. The derivation of the result involves utilizing the equivalence between the square of distance measure and that of the corresponding score.

\begin{lemma}\label{lemma: receive data distance} \emph{Conditioned on the parameters $\{\bw, b\}$ of a binary SVM classifier, channel coefficient $h_k$, and channel noise $\bz_k$, the distance from a received data sample $\hat{\bx}$ to the decision boundary is given as:
\begin{equation}
d({\hat{\bx}})=\sqrt{s^2({{\bf x}})+2s(\bx)\times\frac{\bw^{\sf T}\widetilde{\bf z}_k}{\|\bw\|}+\l(\frac{\bw^{\sf T}\widetilde{\bf z}_k}{\|\bw\|}\r)^2},
\end{equation}
 where $s\l(\cdot\r)$ is the output score given in \eqref{eq: redef score}, and $\widetilde{\bf z}_k=\frac{1}{\sqrt{P}}\Re\l(\frac{{h_k}^*}{\|{h_k}\|^2}{\bf z}_k\r)$ is the equivalently noise after decoding. }
\end{lemma}

According to Lemma~\ref{lemma: receive data distance}, the key step for deriving DII is to find the distribution of the projected channel noise $\frac{\bw^{\sf T}\widetilde{\bf z}_k}{\|\bw\|}$.  The derivation simply involves projecting the high-dimensional Gaussian distribution onto a particular direction specified by $\bold{w}$,  which yields a univariate Gaussian distribution as elaborated below.

\begin{lemma}\label{lemma: projected noise distribution} \emph{Given the specific direction ${\bw}/{\|\bw\|}$, the projected channel noise follows a Gaussian distribution:
\begin{equation}\label{eq:data distribution}
\frac{\bw^{\sf T}\widetilde{\bf z}_k}{\|\bw\|} \sim {\cal N}\l(0 ,\frac{1}{\SNR_k}\r).
\end{equation}
 }
\end{lemma}

Applying the expectation and variance of projected channel noise (see Lemma \ref{lemma: projected noise distribution}) into Lemma~\ref{lemma: receive data distance}, the expected distance of a received data sample is presented in the following lemma.  

\begin{lemma}\label{lemma: expected dis}\emph{The expected distance from the received data sample $\hat{\bf x}_{k,n}$ to the decision boundary~is  
  \begin{align}
  \E_{\bz_k}\l[d^2({\hat{\bf x}_{k,n}})\r]&=d^2({{\bf x}_{k,n}})+\frac{1}{\SNR_k}.
 \end{align}
 }
 \end{lemma}

 \begin{figure}[t]
\begin{center}
{\includegraphics[width=16cm]{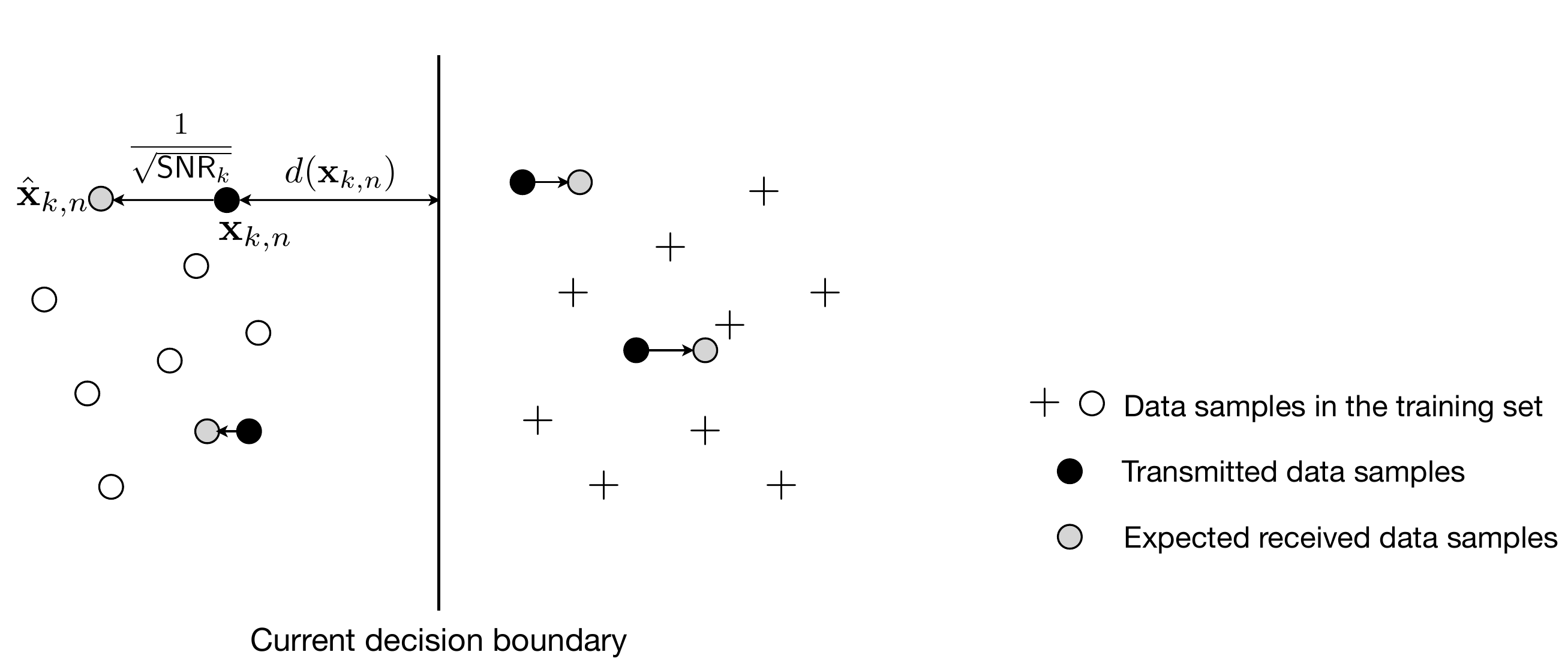}}\\
\vspace{-10pt} 
\caption{Illustration of the derived DII and the effect of channel noise.}
\vspace{-30pt} 
\label{Fig: Unlabeled DII}
\end{center}
\end{figure}

Lemma~\ref{lemma: expected dis} suggests that the channel fading and noise tend to degrade the data importance.  The effect of noise on the receive data importance can be further illustrated in Fig.~\ref{Fig: Unlabeled DII}, where the solid black dots represent the transmitted data samples.  The corresponding received data sample is expected to be the grey dot, which is more likely to be pushed away from the decision boundary, i.e., it suffers importance reduction.  The degradation is proportional to the power of channel noise (or the inverse of SNR).  The result is aligned with the intuition that channel noise is harmful and can not be exploited for improving learning performance.  

With Lemma~\ref{lemma: expected dis}, we are ready to derive a closed-form expression of DII, as shown in the following proposition. 

 \begin{proposition}\label{prop: imp measure without label} \emph{Consider the training of a binary SVM classifier at the edge and the model is broadcast to the edge devices for data uncertainty evaluation.  Given the local dataset $D_k=\{\bx_{k,1},\bx_{k,2},\cdots,\bx_{k,N}\}$ of the $k$-th edge device and  $\SNR_k$, the DII is given as
 \begin{align}\label{eq: imp measure}
I_k= -{\frac{1}{\SNR_k}+\max_{n\in\mathcal{N}} {\mathcal{U}_{\sf d}\l({\bf x}_{k,n}\r)} }\ ,
\end{align}
where $\mathcal{N}=\{1,2,\cdots,N\}$ represents the sample index set, and $\mathcal{U}_{\sf d}\l(\cdot\r)$ is a distance-based uncertainty measure defined in \eqref{eq: dis un}.}
\end{proposition}

\begin{remark}[How does the local buffer size affect DII?]\label{remark: buff size without label}\emph{With the increase of buffer size, the DII is dominated by the SNR due to convergence of the second term related to data uncertainty towards zero.  A larger buffer size suggests potentially higher diversity of the dataset for selection.  Specifically, 
 \begin{align}
 \lim_{N \to \infty}   \max_{n\in\mathcal{N}_k} {\mathcal{U}_{\sf d}\l({\bf x}_{k,n}\r)} =\lim_{N \to \infty}   \min_{n\in\mathcal{N}_k}  \ d^2\l({\bf x}_{k,n}\r) \to 0, \ \forall k.
\end{align}
 As a result, the DII in the case of large buffer becomes:
 \begin{align}
\lim_{N \to \infty} I_k= -\frac{1}{\SNR_k}.
\end{align}
}
\end{remark}
  
The developed DII for binary SVM can be extended to multi-class SVM.  Given a data sample $\bold{x}$, the distance based uncertainty is evaluated based on the predicted label as illustrated in the sequel.  Specifically, a $C$-class SVM classifier is implemented by $L = C(C-1)/2$ \emph{one-versus-one} binary component classifiers that each trained using the samples from the two concerned classes only \cite{platt2000large}.  To predict the label $\hat c$, the output $L$-dimension vector, denoted as  $\bold{s} = [s_1({\bf x}),s_2({\bf x}), \cdots, s_L({\bf x})]$ is compared with a \emph{reference coding matrix} of size $C \times L$, denoted by $\bold{M}$.  An example of the reference coding matrix with $C$ = 4  is provided as follows:
\vspace{-10pt}
\begin{small}
\begin{align*}& \bold{M}  \!\!= \quad
\bordermatrix{%
    & _{\rm  binary 1}      &  _{\rm binary 2}    & _{\rm binary 3}   &_ {\rm binary 4} & _{\rm binary 5} & _{\rm binary 6} 
\cr 
_{{\rm class1}}    & 1         & 1       &1     & 0  & 0  & 0\cr
_{\rm class2}    & -1        & 0       &0     & 1  &1   &0 \cr
_{\rm class3}    &0          &-1      &0      &-1  &0    &1 \cr
_{\rm class4}    & 0         & 0       &-1     &0   &-1   &-1 \cr
},
\end{align*}
\end{small}
where each row gives the ``reference output pattern'' corresponding to the associated class.  Given  $\bold{M}$, the prediction of the class index of $\bold{s}$  involves simply comparing the Hamming distances between $\bold{s}$ and different rows in $\bold{M}$, and choosing the row index with the smallest distance as the predicted class index: 
\begin{align}\label{Hamming_dist}
\hat{c} =  \arg\min_{c} \sum_{\ell=1}^{L}|m_{c\ell}|[1-{\rm sgn}(m_{c\ell}s_\ell({\bf x}))]/2,
\end{align}
 where $m_{c\ell}$ denotes the $\ell$-th element in vector $\bold{m}_c$, and $\rm sgn(x)$ denotes the sign function taking a value from $\{1,0,-1\}$ corresponding to the cases $x>0$, $x=0$ and $x<0$, respectively.  
Having obtained the predicted label $\hat{c}$, the distance based uncertainty is averaged over all the effective component classifiers of the predicted label and DII is defined below:
 \begin{align}\label{eq: DII multiSVM}
I_k= -{\frac{1}{\SNR_k}+\max_{n\in\mathcal{N}}\Big\{-\frac{1}{C-1}\sum_{\ell=1}^{L}|m_{\hat{c}\ell}{s_\ell({\bf x}_{k,n})}|^2 }\Big\}.
\end{align}


\subsection{Importance-Aware Scheduling }

In this section, the importance-aware scheduling is designed for binary SVM classification. Specifically, the edge  sever schedules the device with highest value of DII.  The design can be extended to multi-class SVM following the procedure described in the preceding section. Given the derived DII in Proposition~\ref{prop: imp measure without label}, the resultant scheme is described in Scheme~\ref{Sch:Scheduling} below.  

The summation form of DII in \eqref{eq: scheduling unlabel} elegantly incorporates both data uncertainty and channel quality in the design of scheduling criterion, which provides a simple mechanism for simultaneous exploitation of multiuser data-and-channel diversity.  Any criterion purely exploits only a single type of diversity may compromise the learning performance degradation and lead to inefficient use of radio resources.  Particularly, a scheduling criterion based on only SNR (only exploiting channel diversity, see Scheme~\ref{scheme: channel aware})  is prone to selecting a useless data sample which has little contribution to refining the decision boundary.  On the other hand, the one based on data importance (only exploiting data diversity, see Scheme~\ref{scheme: AL}) may suffer from selecting highly noisy data, and thereby compromise the learning.

\begin{framed}
\vspace{-10pt} 
\begin{scheme}[Importance-aware scheduling for binary SVM]\label{scheme: importance-aware without label} \emph{Consider the acquisition of a data sample from multiple edge devices in an edge learning system.  The edge server schedules device $k^*$  for data transmission if
\begin{equation}\label{eq: scheduling unlabel}
k^*=\arg \max_{k}\  \l\{-\frac{1}{\SNR_k}+\max_{n\in\mathcal{N}_k} {\mathcal{U}_{\sf d}\l({\bf x}_{k,n}\r)} \r\},
\end{equation}
where $\mathcal{U}_{\sf d}(\cdot)$ is the distance based uncertainty defined in \eqref{eq: dis un}.
}
\label{Sch:Scheduling}
\end{scheme}
\vspace{-10pt} 
\end{framed}

 \begin{remark}[How does a transmit SNR affect scheduling?]\label{remark: SNR effect without label}\emph{The effect of the transmit SNR $P/\sigma^2$ on scheduling 
 can be understood by rewriting the scheduling scheme using the definition of SNR given in \eqref{eq: def SNR non comb}:
 \begin{equation}\label{eq: rewrite scheduling}
 k^*=\arg \max_{k}\  \l\{-\frac{\sigma^2}{P}\times\frac{1}{2\|h_k\|^2}+\max_{n\in\mathcal{N}_k} {\mathcal{U}_{\sf d}\l({\bf x}_{k,n}\r)} \r\}.
 \end{equation}
 {\color{black}One can observe that the transmit SNR $P/\sigma^2$ is a weight factor to balance the influences of channel quality and data uncertainty on the scheduling decision.  The scheduling schemes for low and high transmit SNR scenarios are discussed as follows.
 \begin{itemize}
  \item \emph{Low transmit SNR:}  This case, wireless channels are unreliable.  The channel diversity is more critical to be exploited for reliably receiving a data sample.  Otherwise, received data samples are severely corrupted by noise and become useless regardless of their uncertainty (importance) prior to transmission.   This fact causes the proposed scheme to enforce a large weight factor (low transmit SNR) for channel quality in the scheduling metric.  Moreover, the scheme is reduced to channel-aware scheduling (see Scheme~\ref{scheme: channel aware}) when the transmit SNR approaches zero. 
  \item \emph{High transmit SNR:} On the contrary, when the transmit SNR is high, it is more critical to exploit the data diversity as all wireless data channels are reliable.  In this case, acquiring data samples with high original uncertainty values accelerates the model training.  This fact is translated into the small weight factor (high transmit SNR)  for channel quality so as to make data uncertainty dominant in scheduling decision making.  If the transmit SNR is sufficiently large, the scheduling scheme reduces to pure important data selection (named data-aware scheduling in Scheme~\ref{scheme: AL}) as the first term in \eqref{eq: rewrite scheduling} vanishes.   
  \end{itemize}
  }}
\end{remark}
 
Last, the two mentioned conventional schemes that are special cases of importance-aware scheduling are presented as follows.

\begin{framed}
\vspace{-10pt} 
\begin{scheme}[Channel-aware scheduling]\label{scheme: channel aware} \emph{Consider the acquisition of a data sample from multiple edge devices in an edge learning system.  The edge server schedules the $k^*$  device for data transmission if
\begin{equation} \label{eq: max SNR:a} 
k^*=\arg \max_k\  \SNR_k,
\end{equation}
where ${ \SNR}_k$ is defined in \eqref{eq: def SNR non comb}, and the transmitted data sample is randomly selected from the scheduled edge device.
}
\end{scheme}
\vspace{-10pt} 
\end{framed}

\begin{framed}
\vspace{-10pt} 
\begin{scheme}[Data-aware scheduling]\label{scheme: AL} \emph{Consider the acquisition of a data sample from multiple edge devices in an edge learning system.  The edge server schedules the $k^*$  device for data transmission if
\begin{equation} \label{eq: max SNR:b}
k^*=\arg \max_k\  \max_{n\in\mathcal{N}_k} {\mathcal{U}_{\sf d}\l({\bf x}_{k,n}\r)} ,
\end{equation}
where $\mathcal{U}_{\sf d}(\cdot)$ is the distanced based uncertainty defined in \eqref{eq: dis un}.}
\end{scheme}
\vspace{-10pt} 
\end{framed}


\subsection{Extension to General Classifier Models}\label{sec: general classifiers}
In this section, the proposed importance-aware scheduling targeting for SVM classifier is extended to a general model.  However, the derivation for SVM may not be directly applied to a generic classifier (e.g., CNN), due to the lack of an explicitly defined functional mapping from input noisy data to the output score.  Nevertheless, the general form of the DII derived in the SVM setting that cascades the data reliability and data importance measures in a summation form is applicable to a generic model.  This motivates the simple generalization of the importance-aware scheduling by replacing the uncertainty measure in \eqref{eq: dis un} targeting SVM with a general measure, which can be properly chosen depending the specific learning model. The modified scheme is descried as follows.  

\begin{framed}
\vspace{-10pt} 
\begin{scheme}[Importance-aware scheduling for a generic classifier]\label{scheme: generic importance-aware without label} \emph{Consider the acquisition of a data sample from multiple edge devices in an edge learning system.  The edge server schedules the $k^*$  device for data transmission if
\begin{equation}
k^*=\arg \max_{k}\  \l\{-\frac{1}{\SNR_k}+\max_{n\in\mathcal{N}_k}  {\mathcal{U}_{\sf x}\l({\bf x}_{k,n}\r)} \r\},
\end{equation}
where $\mathcal{U}_{\sf x}$  is a general uncertainty measure.  In particular, it can be the entropy $ \mathcal{U}_{\sf e}$  defined in \eqref{eq: entropy un} if CNN classifier model is adopted.  
}
\end{scheme}
\vspace{-10pt} 
\end{framed}
As a guideline, the selection of the uncertainty measure should allow easy computation using the model output.  For example, for SVM, the output score evaluated using the linear decision boundaries allows easy evaluation of the distance-based uncertainty.  On the other hand, for CNN, the softmax output, which gives the posterior probability for each predicted class, makes the entropy to be a more natural choice for measuring uncertainty.

\section{Implementation Issues and Solutions} \label{sec: discussions}
{ \color{black}
\subsection{Importance-Aware Scheduling With Label Information}
 In the preceding sections, the data importance (uncertainty) is evaluated and a scheduling decision made based on unlabelled data with a label generated after scheduling.  This targets the scenario where labelling (e.g., by a human labeler) is costly.  However, in some cases, data at the edge devices are generated together with labels.  For example, training data for auto-driving (e.g., outputs of cameras and radar) are automatically labelled by sensing the driving decisions of a human driver.  Then, the design of DII should exploit label information to further accelerate the learning speed. 
 
 \begin{figure}[t]
\begin{center}
{\includegraphics[width=10cm]{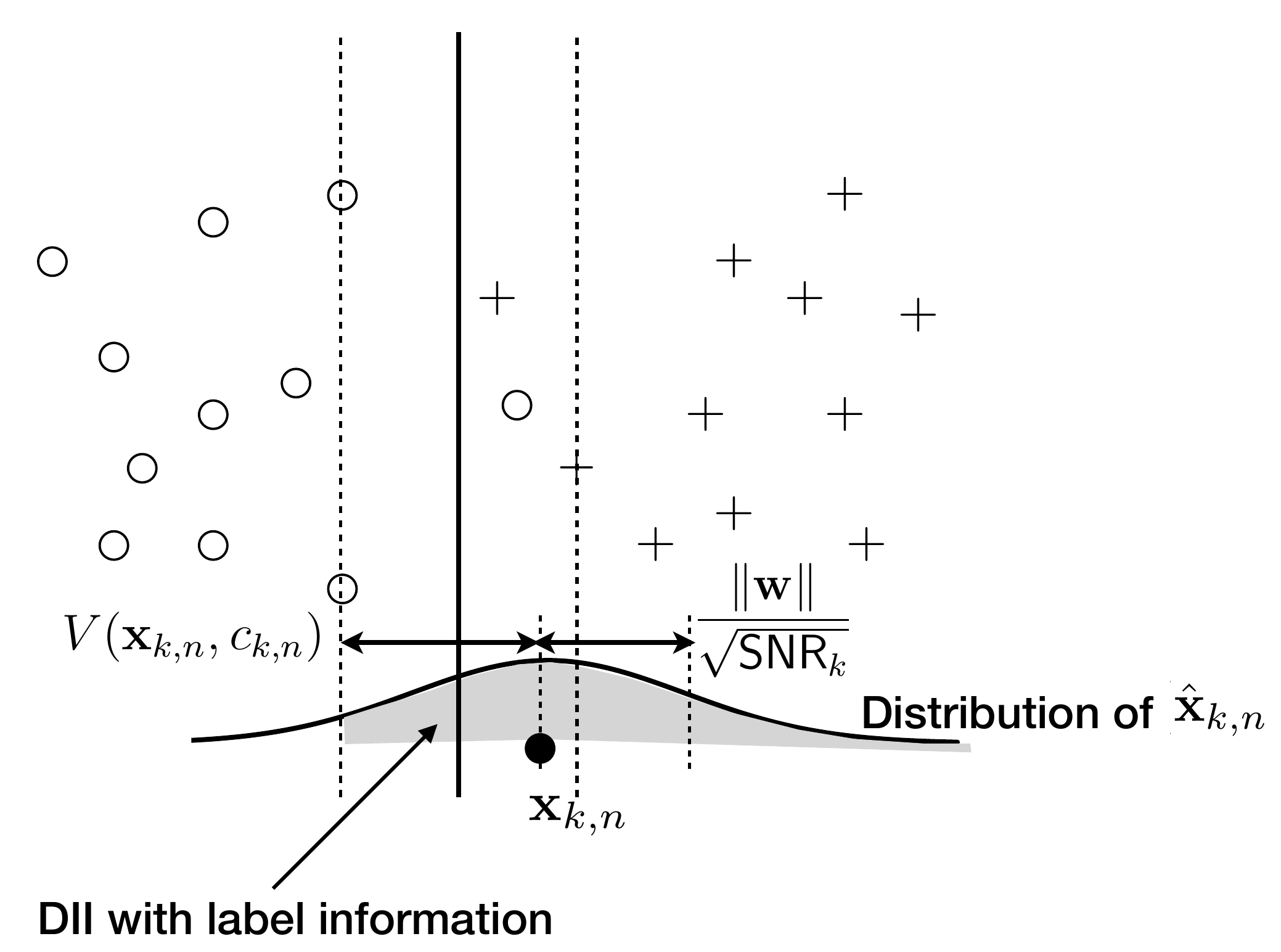}}\\
\caption{Illustration of the DII with label inforamtion.}
\vspace{-20pt} 
\label{Fig: ProbSV}
\end{center}
\end{figure}
 
The design of DII for the current case of all labelled data is based on the fact in the incremental learning of SVM that a newly added data sample to the training dataset can update the improve model if it is a support vector \cite{cauwenberghs2001incremental}.  As a result, the DII is defined to be the expectation of model update.  With the definition of support vector given in \eqref{eq: sv}, the event of model update, denoted as $\mathcal{V}$, is defined as that both transmitted and received data samples are support vectors:   $\{\mathcal{V}(\hat{\bx}|\bx, c)|\l(V(\bx,c)\geq0\r) \cap \l(V(\hat{\bx},c)\geq0\r) \}$.  The event ensures the model update is due to the important data sample instead of the channel noise.  Mathematically, the DII with label information is defined as follows:  
\begin{align}
I_k&=\max_{n\in\mathcal{N}}\E_{\bz_k}\l[\mathcal{V}(\hat{\bx}_{k,n}|\bx_{k,n}, c_{k,n})\r] \\ 
&=\max_{n\in\mathcal{N}}\int_{-V(\bx_{k,n},c_{k,n})}^{\infty} \sqrt{\frac{\SNR_k}{2 \pi  \|\bw\|^2} }\exp\l(-\frac{t^2}{2 \|\bw\|^2/{\SNR_k} }\r)dt \nn\\
&=\max_{n\in\mathcal{N}}\ \frac{1}{2}\l[1+\ {\rm erf}\l(\frac{V(\bx_{k,n},c_{k,n})}{\sqrt{{2\|{\bf w}\|^2}/{\SNR_k}}}\r) \r] , \  \forall \ V(\bx_{k,n},c_{k,n})\geq0 , \label{eq: ratio SNR loss}
\end{align} 
where $V(\bx,c)$ has been given in \eqref{eq: sv}.  The result is graphically shown in Fig.~\ref{Fig: ProbSV} as the shaded area.  One can notice that DII is a probability that requires the variation of channel noise lies inside the margin boundary, which is determined by the ratio of $V(\bx,c)$ and noise power $\frac{1}{\sqrt{\SNR_k}}$ as derived in \eqref{eq: ratio SNR loss}. Then the importance-aware scheduling with label information is designed as follows based on the simplified DII. 

\begin{framed}
\vspace{-10pt} 
\begin{scheme}[Importance-aware scheduling with label information]\label{scheme: label scheduling} \emph{Consider the acquisition of a data sample from multiple edge devices in an edge learning system.  The edge server schedules the $k^*$  device for data transmission if
\begin{equation}
k^*=\arg \max_{k}\  \l\{\sqrt{\SNR_k}\times\max_{n\in\mathcal{N}_k}  \ \max\big[0,V(\bx_{k,n},c_{k,n})\big]\r\},
\end{equation}
where $V(\bx_{k,n},c_{k,n})$  is defined in \eqref{eq: sv} and $\max\big[0,V(\bx_{k,n},c_{k,n})\big]$ is to pre-select the support vector $(V(\bx_{k,n},c_{k,n})\geq0)$ for transmission.   
}
\end{scheme}
\vspace{-10pt} 
\end{framed}

It is remarked that $\max\big[0,V(\bx_{k,n},c_{k,n})\big]$ is exactly the same as the definition of hinge loss \cite{bishop2006pattern}, and thus DII with label information elegantly incorporates two metrics from communication and learning perspectives.  Compared with the data selection by using the uncertainty measure (without label information), hinge loss is another way to exploit data diversity which has its pros and cons.  
For the data selection based on uncertainty, the new coming data sample near to the decision boundary helps to refine the optimal classifier (reduce the hypothesis space) in a binary search manner.  With the label information, the selected data sample based on hinge loss could cross the decision boundary with a wrong predicted label, thereby achieve a faster rate than the binary search so as to accelerate learning speed.  On the other hand, when the hypothesis space is small, the  hinge loss may guide to select the outlier in  non-separable dataset, and thus mislead the classifier to the opposite side.   
}

%
%

\vspace{-10pt}

\subsection{Compressed Model for DII Evaluation}\label{subsec: compressed model}

 \begin{figure}[t]
\begin{center}
{\includegraphics[width=15cm]{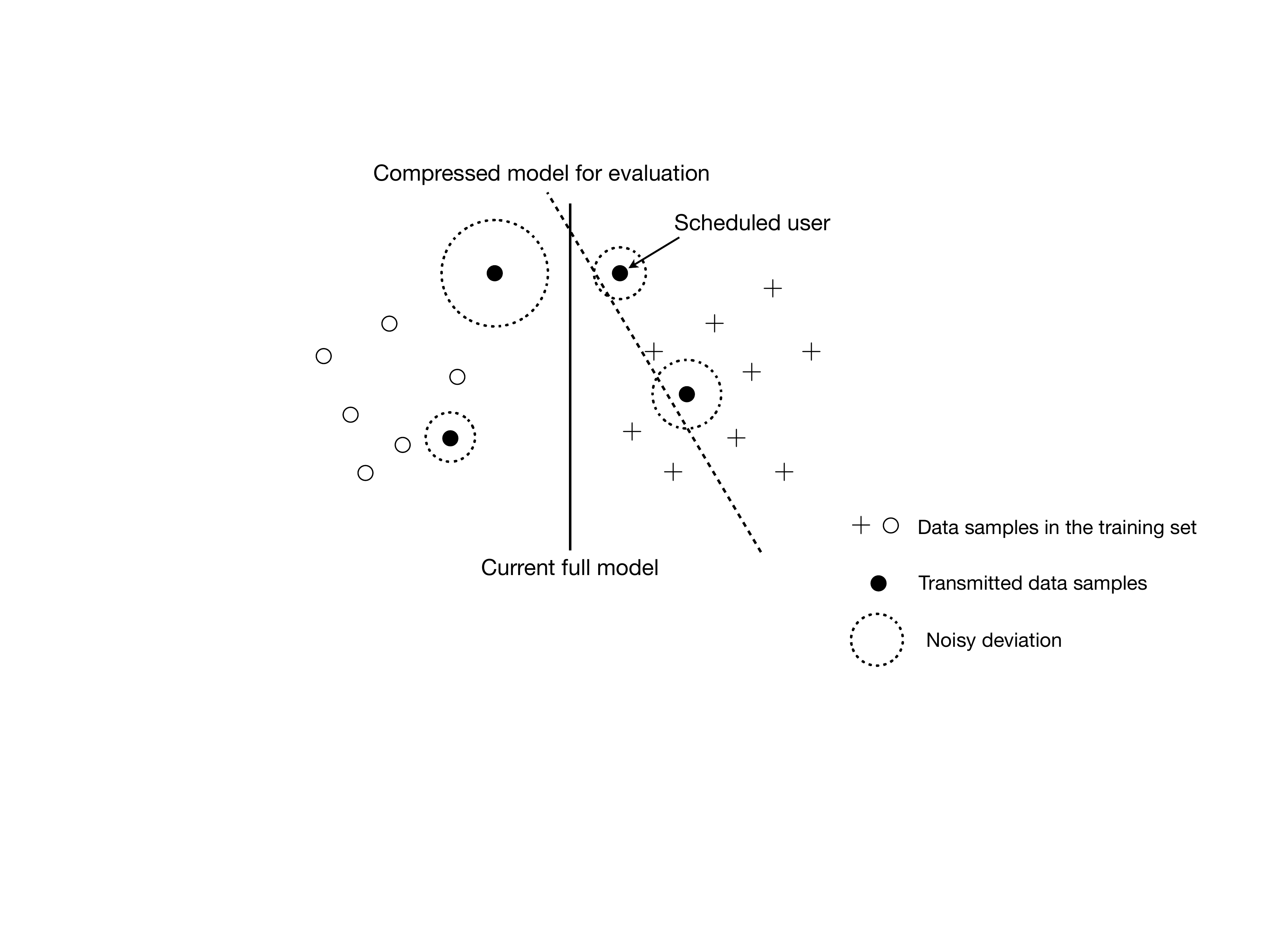}}\\
\caption{Effect of the compressed model for unlabeled DII evaluation.}
\vspace{-30pt} 
\label{Fig: CompressModel}
\end{center}
\end{figure}

One practical issue in implementing importance-aware scheduling is high local computing complexity of DII evaluation.  Specifically, the energy consumption and the requirement of computing resources for data uncertainty evaluation using the full model may be too costly for a resource constrained edge device. This motivates the use of compressed model for DII evaluation that can reduce local computing complexity without significantly compromising the scheduling performance. Underpinning the design is the fact that no performance loss will be incurred as long as the compressed model can provide sufficient differentiability amongst data in terms of DII values.  To illustrate this point, as shown  in Fig.~\ref{Fig: CompressModel}, the scheduling based on the DII evaluated using the compressed model may make exactly the same decision as that based on DII evaluated using the full model.  This fact indicates the existence of model redundancy to be reduced for efficient data uncertainty evaluation.  
To avoid performance degradation, the compression ratio should be properly selected according to data distribution, SNR and the number of users. For example, a highly separable dataset and sparse device population allows large space for model compression.  On the other hand, for the low transmit SNR scenario, the scheduling scheme relies less on data importance, thus requires less accurate model to evaluate its value.

In the experiment, we vary the compression ratio $C_r \in (0,1)$, representing the ratio of parameters remained, and show its effect on learning performance. 
{\color{black}Given the compression ratio, we could further characterize the computational efficiency of data uncertainty evaluation at edge devices, which is determined by the number of computer operations for calculating the output score.  In general, it is scaled by the total number of parameters $W$ in the learning model. Specifically, for a linear classifier like SVM, the overall local computational cost is $\mathcal{O}\l(WC_rN\r)$, for calculating linear score functions of $N$ data samples.   On the other hand, computing an output score of CNN requires forward propagation in the compressed neural network,  and thus the local computational cost is also  $\mathcal{O}\l(WC_rN\r)$ as indicated in \cite{bishop2006pattern}. }

\subsection{Several Methods for  Performance Enhancement}\label{sec: mobility and update}
As the importance-aware scheduling exploits data diversity, the resultant performance highly depends on total distributed data samples for selection, which is proportional to the number of users, local buffer size, the update frequency of local buffer and user mobility as characterized in the following.  Assume the number of edge users is $K$ in the network in each time slot,  during $T$ slots of the wireless data acquisition, the  available number of data samples for scheduling is given as 
\begin{equation}\label{eq:data diversity}
\mathcal{D}(T)=KN+\sum_{t=2}^{T}\l[K-P_u(t)\r]P_d(t)+P_u(t)N,
\end{equation}
where $P_d(t) \in \{0, 1, 2, \cdots, N\}$ is the number of updated samples in each devices, and $P_u (t)\in \{ 0, 1, 2, \cdots, K\}$ is the number of users replaced in the coverage cell due to mobility.  The result in \eqref{eq:data diversity} suggests that the performance of importance-aware scheduling can be improved in several possible ways such as increasing the number of users $K$ and the buffer size $N$, updating the local buffered samples with a higher frequency, or utilizing user mobility. They are verified by simulation to be effective.

%
%
%
%

\vspace{-10pt}
\section{Experimental Results}\label{sec:simulation}

\vspace{-10pt}
\subsection{Experiment Setup}
The default experiment settings are  as follows unless specified otherwise.  The number of edge devices is $K=10$.  Each device is equipped with a local buffer, with the size $N=10$, and updates one of outdated data samples with a new one, denoted as $P_d(t)=1$, for an arbitrary slot $t$.
Consider the static user case where the number of users replaced in the coverage cell is set as $P_u(t)=0$ for all $t$. The maximum transmission budget $T$ for the learning task is given as $100$ and $1,000$ (channel uses) for binary and multi-class classifications, respectively.  Under the transmission budget constraint, we consider the test accuracy as the performance metric. All results are averaged over 150 and 20 experiments for binary SVM and multi-class CNN, respectively.

\subsubsection{Channel Model} We assume the classic Rayleigh fading channel with channel coefficients $\{h_k\}$ following i.i.d. complex Gaussian distribution ${\cal CN}(0,1)$. The average transmit SNR defined as {\color{black}$\bar{\rho}=P/\sigma^2$ is by default set as $15$ dB.

\subsubsection{Experimental Dataset} We consider the learning task of training a classifier using the well-known MNIST dataset of handwritten digits.   
The training and test sets consist of  $60,000$ and $10,000$ samples, respectively.  Each sample is a grey-valued image of $28 \times 28$ pixels that gives the sample dimensions $p=784$.  
 The experiments of multi-class classification are conducted by using the whole dataset.
For binary classification, we choose the relatively less differentiable class pair of  ``3" and ``5" (according to \emph{t-distributed stochastic neighbor embedding} visualization) from the whole data set, including $11,552$ training samples and $1902$ test samples. 
The training set used in experiments is partitioned as follows.  At the edge server, the initially available training dataset ${\cal L}_0$ is constructed by randomly sampling $2$ data samples for each class.  The remaining training data are evenly and randomly partitioned for constructing the local datasets at edge devices, which are used for updating local buffers.  The placement of data samples into a local buffer are randomly sampled from its local dataset, following the update rules as discussed at the beginning of this section.    


\subsubsection{Learning Model Implementation} The considered classifier models include binary SVM and CNN.  For binary SVM,  the soft-margin SVM is implemented with slack variable set as $1$.   \emph{Iterative Single Data Algorithm} (ISDA) \cite{kecman2005iterative} is used for solving the SVM problem with maximum $10^6$ iterations.  For the implementation of CNN, we use a  $6$-layer CNN  including two $3 \times3$ convolution layers with batch normalization before ReLu activation (the first with $16$ channels, the second with $32$), the first one followed with a $2\times2$ max pooling layer and the second one followed with a fully connected layer, a softmax layer, and a final classification layer.   The model is trained using stochastic gradient descent with momentum~\cite{sutskever2013importance}. The mini-batch size is $2048$, and the number of epochs is $120$.  To accelerate training, the CNN is updated in a batch mode with the incremental sample size set as $10$.  The broadcast model is uncompressed, i.e., $C_r=1$, by default.

\subsection{Learning Performance for SVM}

\subsubsection{Convergence Rate} 

\begin{figure}[t]
\begin{center}
{\includegraphics[width=8cm]{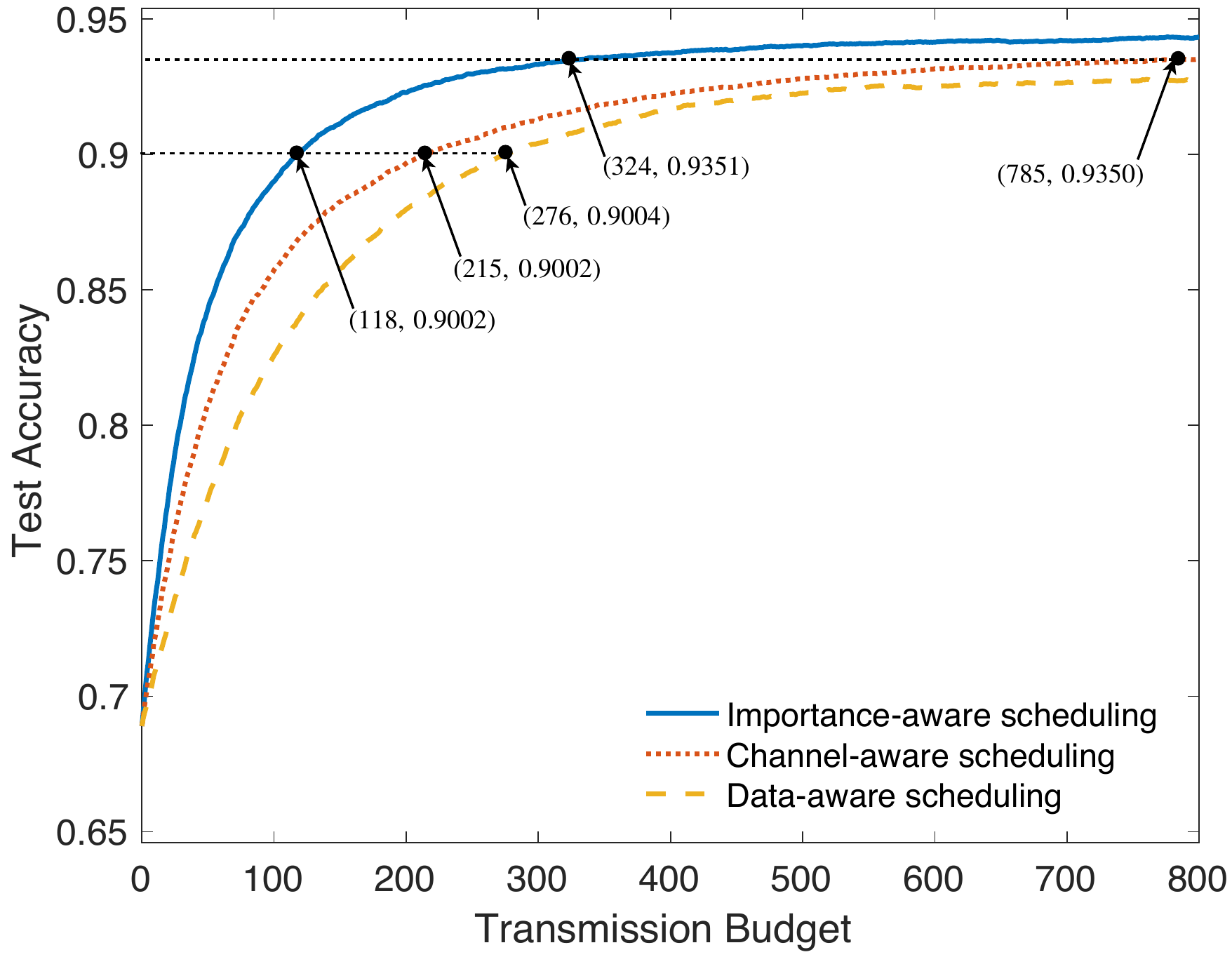}}\vspace{-15pt}\\
\caption{Test accuracy versus transmission budget.}
\vspace{-30pt} 
\label{Fig: budget}
\end{center}
\end{figure}

In Fig.~\ref{Fig: budget}, the learning performance of the proposed importance-aware scheduling is compared with two baseline schemes, namely the channel-aware scheduling and data-aware scheduling, corresponding to pure channel selection and important data selection respectively. The transmission budget varies from $0$ to its maximum value set as $800$ for ensuring model convergence of all schemes. It is observed that the proposed scheme outperforms the two benchmarks. Specifically, if the targeted accuracy is $0.9$, the required budget is $118$ for importance-aware scheduling while it is $215$ and $276$ for channel-aware scheduling and data-aware scheduling respectively.  Thus, it saves more than half budget to achieve the targeted performance by using importance-aware scheduling. The comparison is more remarkable if targeted accuracy is $0.935$, where the budget requirements are $324$ and $785$ for the proposed scheme and conventional channel-aware scheme respectively. In contrast, data-aware scheduling can not achieve that targeted accuracy within the maximum transmission budget.
This confirms the fast convergence by exploiting both data and channel diversities, and verifies the effectiveness of the proposed scheme for fast edge learning.  In the following experiments, we fix the transmission budget of all schemes and compare their test accuracies instead, which is equivalently to reflect the difference in terms of convergence~rate.

\subsubsection{Multi-user Diversity}

\begin{figure}[t]
\begin{center}
{\includegraphics[width=8cm]{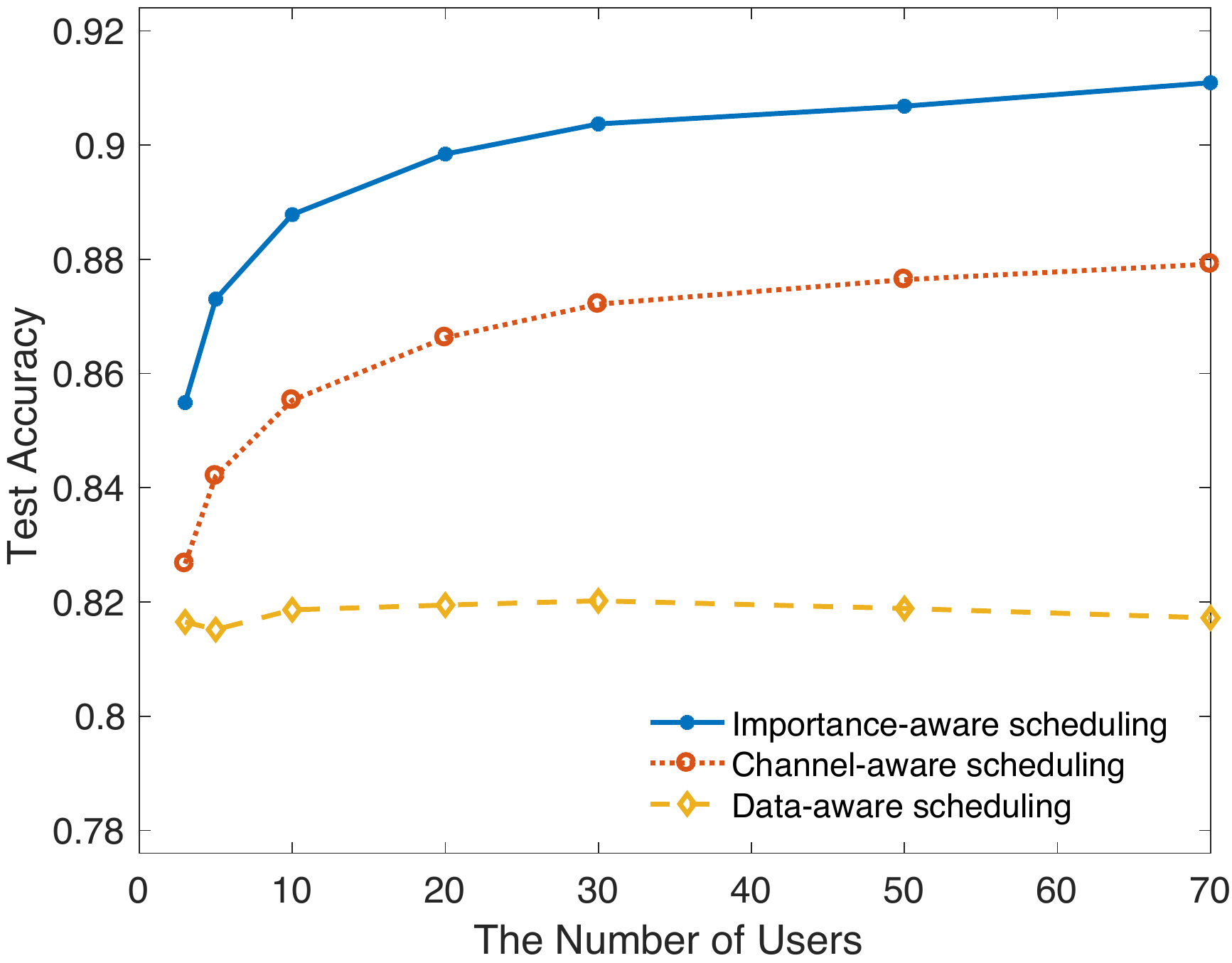}}\vspace{-15pt}\\
\caption{Test accuracy versus the numbers of users.}
\label{Fig: NumUser}
\end{center}
\vspace{-30pt}
\end{figure}

In Fig.~\ref{Fig: NumUser}, we investigate the gain of multi-user diversity by plotting test accuracy over the number of users.  The performance of importance-aware scheduling consistently outperforms two benchmarks in varying number of users scenarios.  This verifies the performance gain by intelligently allocating radio resources according to both data importance and channel condition.  It is observed that the data-aware scheduling is hardly to exploit data diversity in the wireless edge learning scenario.  As the scheme itself is unconscious of channel conditions, the important data samples are contaminated by large channel noise that impeding the performance improvement by data selection.  The result indicates that reliable transmission is the  principle requirement before exploiting data importance for learning.  
Second, by involving more users in the learning system, importance-aware scheduling  achieves larger performance improvement than that of the channel-aware scheduling at the initial stage (e.g., $K<10$). The reason is that  small number of users incurs data deficiency that could be overcome by adding more users.  In contrast, when the number of users is large (e.g., $K>10$), the improvement rates of two schemes are comparable.  In this case, the improvement is mainly due to channel diversity while the gain by exploiting data is saturated if more users are involved.  
 The result reflects that multi-user diversity include two folds, data and channel, which should be jointly exploited for improving learning performance.  The  schemes only exploiting one aspect like the baseline schemes will cause a potential degradation in learning performance.

\subsubsection{Transmit SNR}

\begin{figure}[t]
\begin{center}
{\includegraphics[width=8cm]{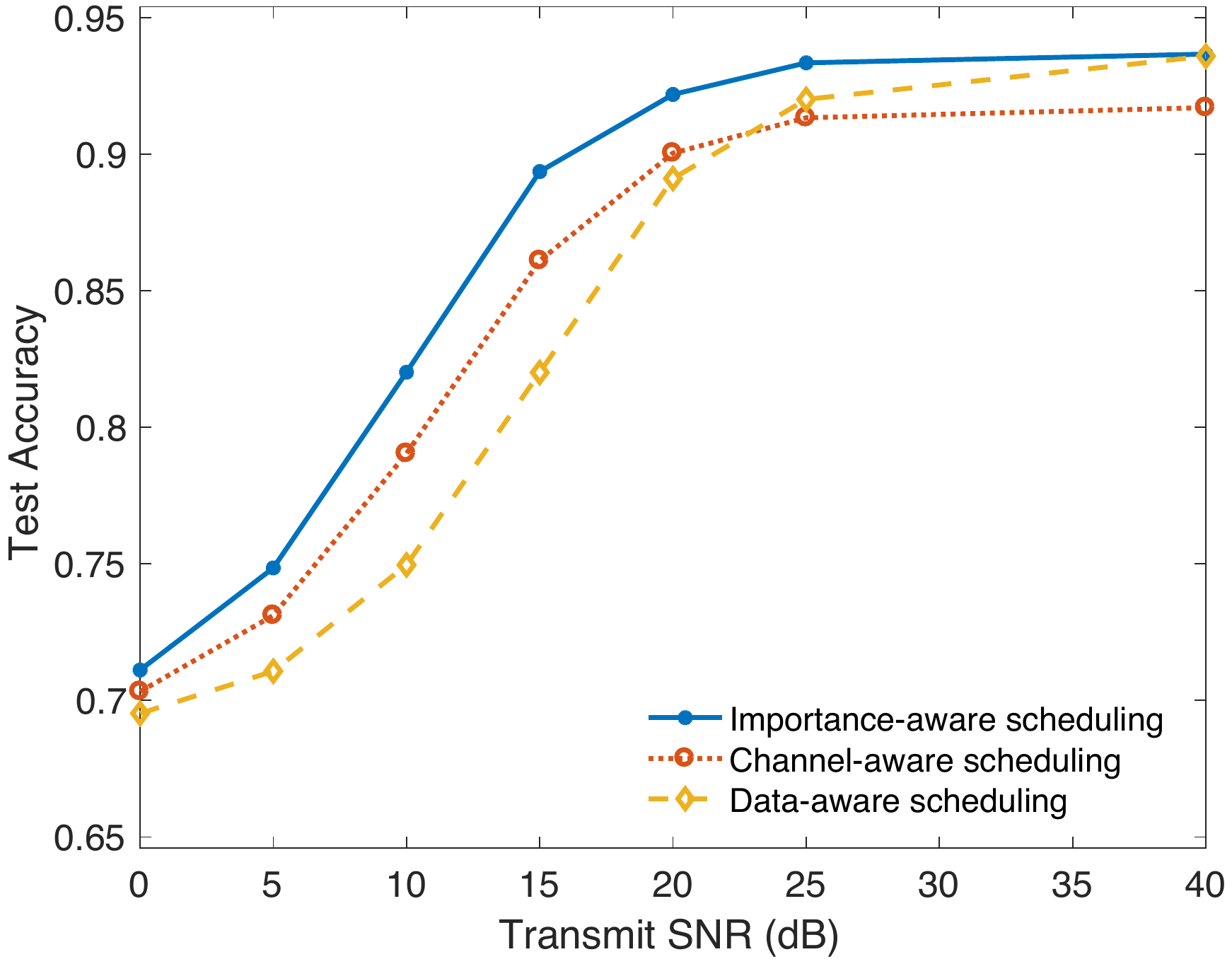}}\vspace{-15pt}\\
\caption{Test accuracy versus the average transmit SNR~$\bar{\rho}$.}
\label{Fig: TransmitSNR}
\end{center}
\vspace{-30pt}
\end{figure}

To demonstrate its robustness against the hostile channel conditions, the proposed importance-aware scheduling is tested under different values of transmit SNR and the results are shown in Fig.~\ref{Fig: TransmitSNR}. 
One can notice that the test accuracy of proposed scheme is always better than that of the two baseline schemes.  The results further substantiates 
the performance gain by jointly exploiting both data and channel diversities.  
To be specific, it is more essential to balance the tradeoff between data importance and data reliability in a moderate transmit SNR  scenario (e.g., $\rho=5 -25\ {\rm dB}$), since the proposed scheme is shown to achieve a more remarkable performance gain than the two benchmarking schemes.  
In contrast, the proposed scheme reduces to channel-aware scheduling and data-aware scheduling in low transmit SNR (e.g., $\rho=0 \ {\rm dB}$) and high transmit SNR (e.g., $\rho=25 \ {\rm dB}$) scenarios respectively, which verifies the discussion in Remark~\ref{remark: SNR effect without label}.  The comparison of three schemes reflects that data reliability is the most critical requirement, since all of schemes suffer severe performance degradations in low SNR scenario. Upon a certain guarantee on data reliability, then the performance can be further improved by exploiting data importance


\subsubsection{Data Diversity}

\begin{figure}[t]
\centering
\subfigure[Impact of buffer size.]{
\label{Fig: BufferSize}
\includegraphics[width=8cm]{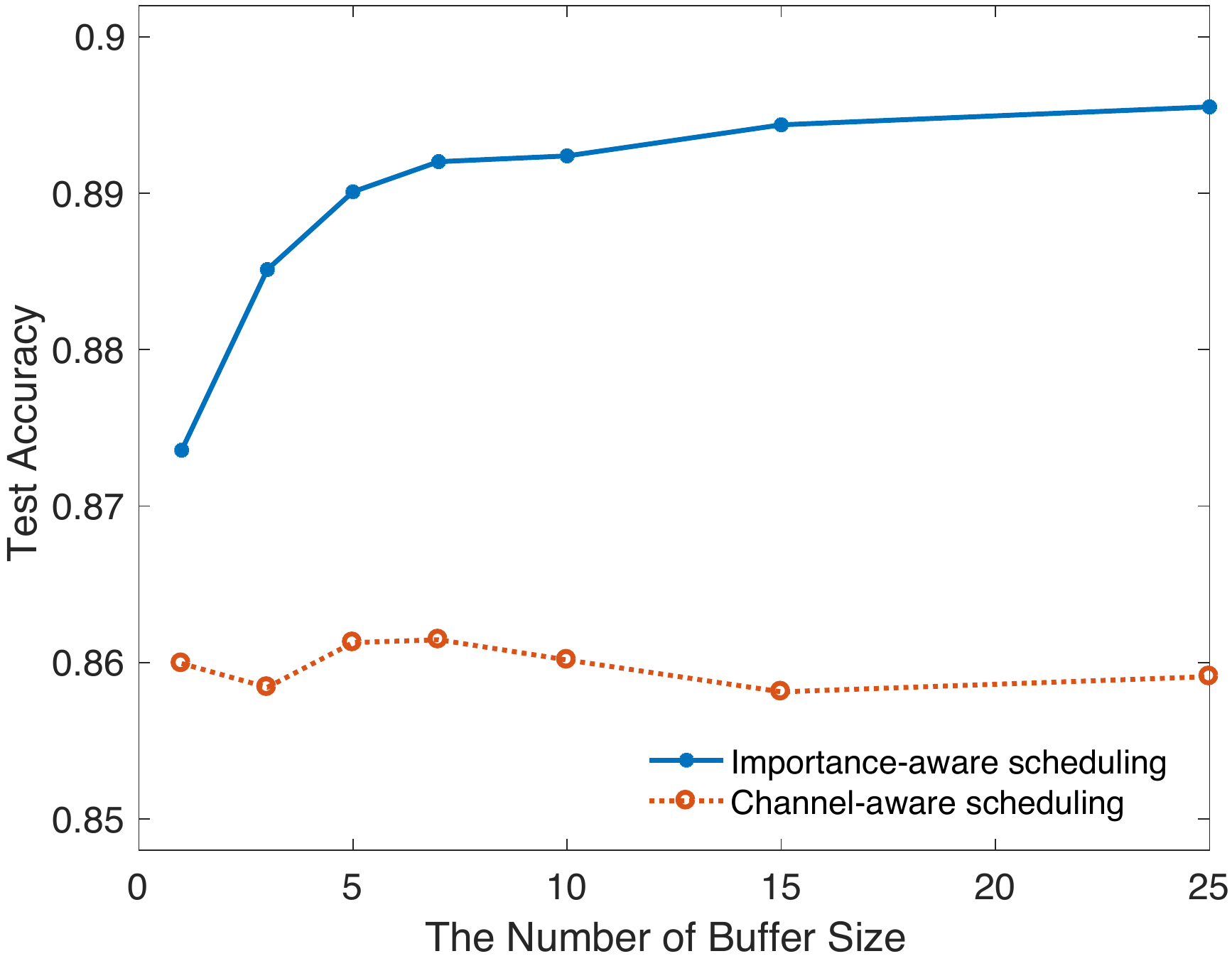}}
\subfigure[Impact of sample-update frequency.]{
\label{Fig: UpdateBuff}
\includegraphics[width=8cm]{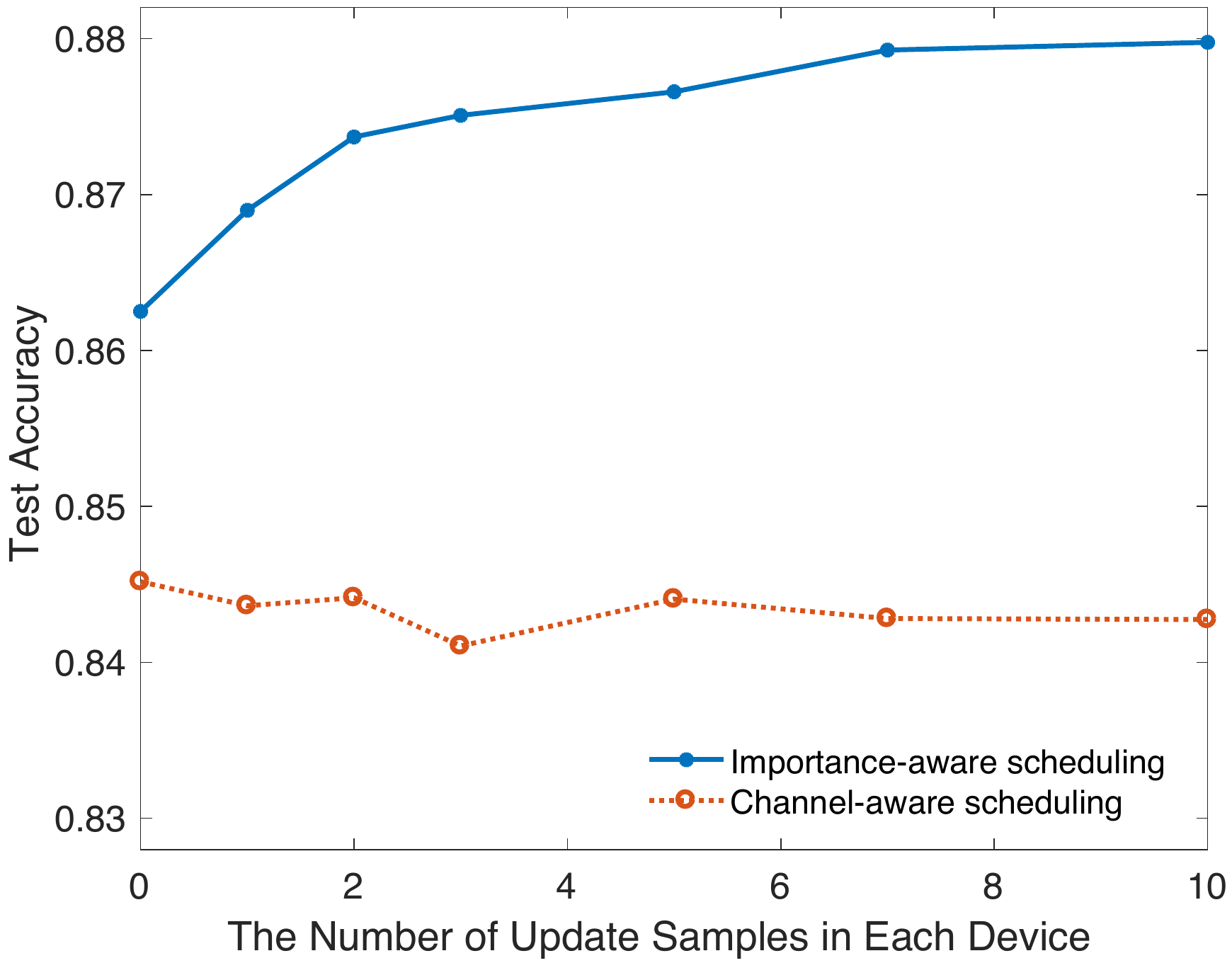}}
\subfigure[Impact of user mobility.]{
\label{Fig: UserMobility}
\includegraphics[width=8cm]{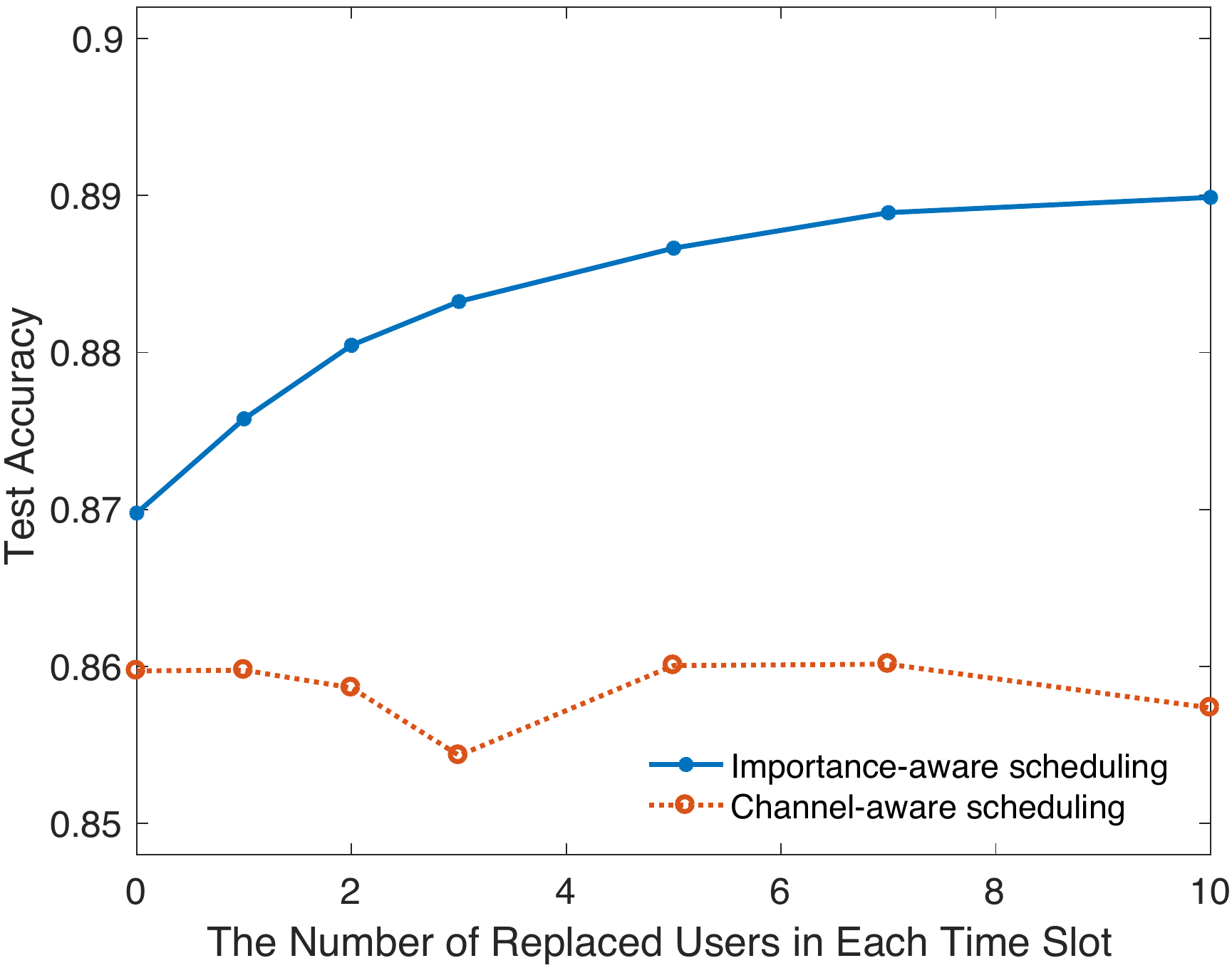}}
\caption{Impact of different approaches for data diversity enhancement. }
\label{Fig: data diversity}
 \vspace{-15pt} 
\end{figure}

Fig.~\ref{Fig: data diversity} demonstrates the performance improvement by increasing the number of available data samples for selection, which depends on the number of users, local buffer size, the update frequency of local buffer and user mobility, as discussed in Section~\ref{sec: mobility and update}.  Since the effect of the number of users has been discussed, this part will focus on the other three, which purely exploit the diversity in distributed data. The relevant results are shown in subfigures respectively, where the performance of importance-aware scheduling is compared with the channel-aware scheduling.  The three figures verify that the conventional scheme is unable to exploit the distributed data samples.  
\begin{itemize}
\item \emph{Local buffer size:}  Fig.~\ref{Fig: BufferSize} presents the performance improvement by increasing the number of buffer size.   The incremental buffers size leads to a remarkable performance improvement at the initial stage ($N<5$), corresponding to a data deficiency scenario. Then the improvement will be saturated if continuously increase the buffer size ($N>15$).  
\item \emph{Update frequency of local buffer:}  Fig.~\ref{Fig: UpdateBuff} displays the performance curves of test accuracy versus the number of update samples in each device.  Note that the update data frequency of local buffer could affect the learning performance only in a data deficiency scenario and the number of users is set as $K=5$ in this experiment.  The result of importance-aware scheduling reveals that the data deficiency can be overcome by frequently updating the samples in local buffer.  
\item \emph{User Mobility:} The user mobility could be specified as the number of replaced users $P_u(t)$ in each time slot, and high mobility corresponds to the large value of $P_u(t)$. In Fig.~\ref{Fig: UserMobility}, the test accuracy is plotted over the number of replaced users.  In this experiment, the buffer size is set as $N=5$ for a data deficiency scenario, and $P_d(t)$ is set as $0$ to reflect the unique performance improvement by exploiting user mobility, which is shown to be prominent.  



\end{itemize}

\subsubsection{Compressed Model for Importance Evaluation}

\begin{figure}[t]
\begin{center}
{\includegraphics[width=8cm]{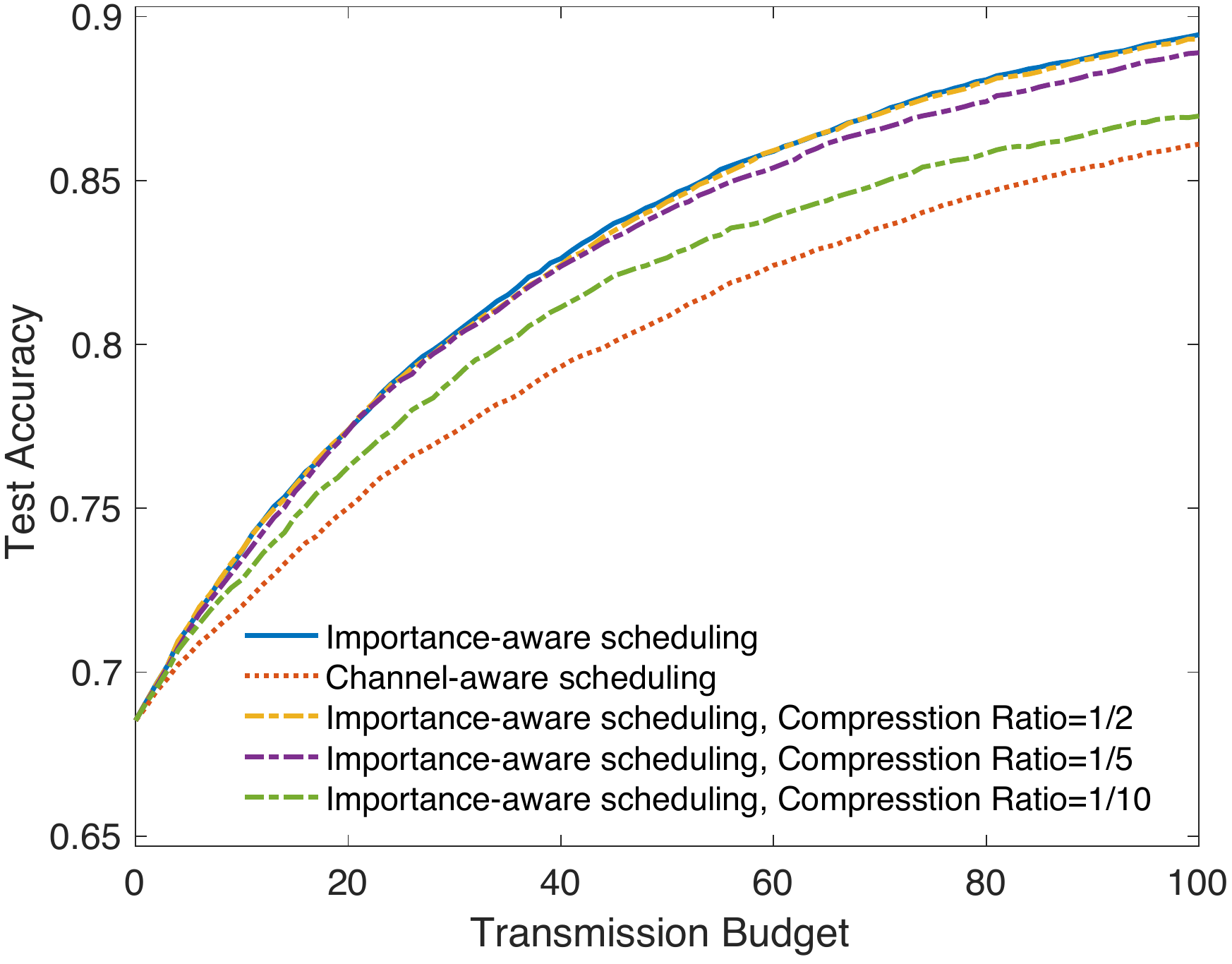}}\vspace{-20pt}\\
\caption{Test accuracy for different evaluation models by varying its compression ratio $C_r$.}
\label{Fig: ModelCompression}
\end{center}
 \vspace{-30pt} 
\end{figure}

In Fig~\ref{Fig: ModelCompression}, the proposed scheme is tested under different importance-evaluation models, by varying the model compression ratio. Although a simple model as SVM, it is able to reduce half computing operations ($C_r=1/2$) without incurring performance loss.  The performance of importance-aware scheduling by using a compressed importance-evaluation model is shown to consistently outperform the channel-aware scheduling, even the number of  model parameters is reduced  by $10 \times$ ($C_r=1/10$).  On the other hand, the performance loss due to model compression is related with the training stage: 
the precision of evaluation model should be increased as the learnt model becomes more accurate, that is, allowing fewer number of model parameters to be reduced.   As shown by the curve of $C_r=1/5$,  at the initial stage, it achieves same performance as the one using uncompressed model, while the performance loss increases as the model being more accurate.  

\subsubsection{Scheduling With Label Inforamtion}

\begin{figure}[t]
\begin{center}
{\includegraphics[width=8cm]{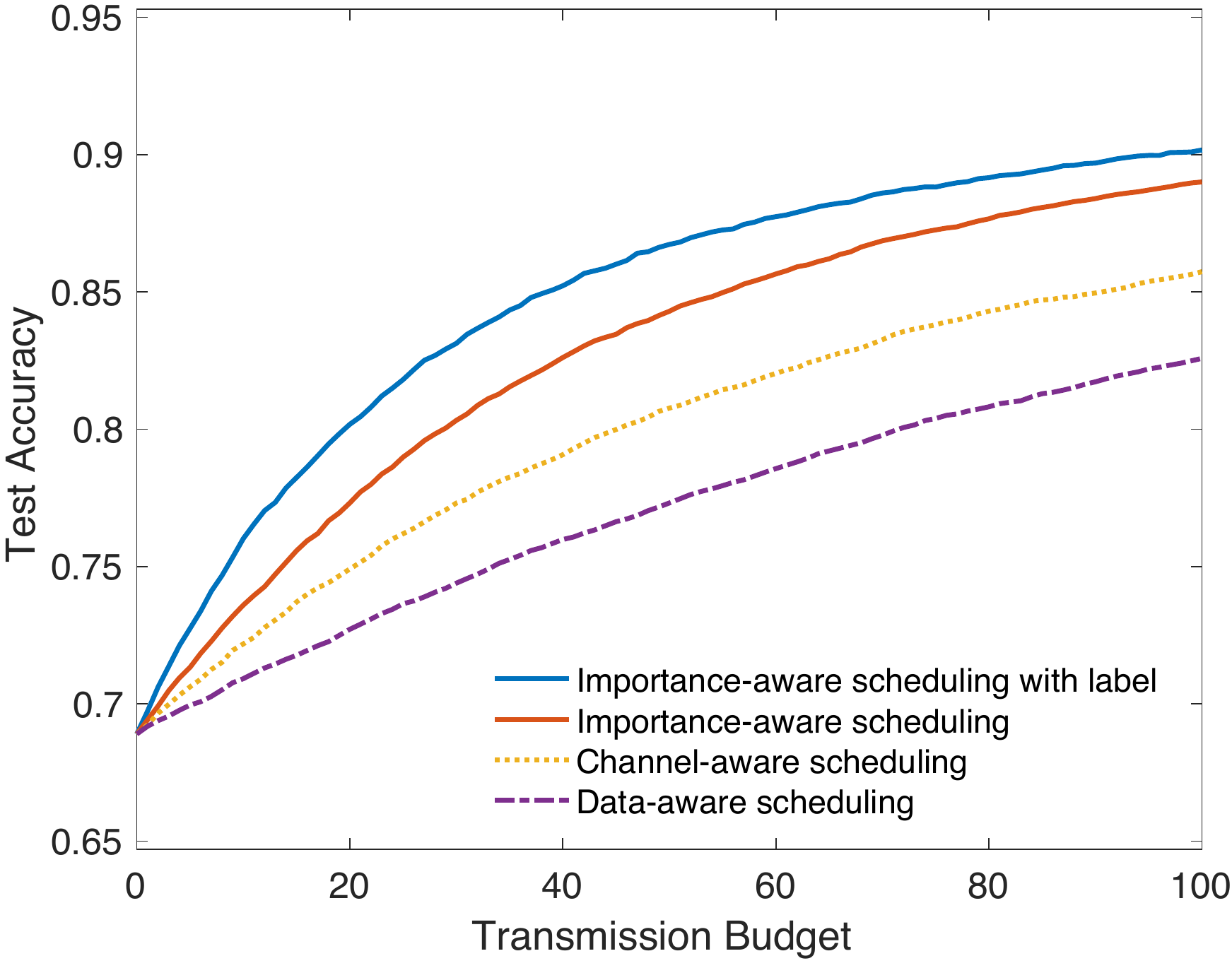}}\vspace{-10pt}\\
\caption{Learning performance evaluation for importance-aware scheduling with label inforamtion.}
\label{Fig: label scheduling}
\end{center}
\vspace{-20pt} 
\end{figure}
{\color{black}
In Fig~\ref{Fig: label scheduling},  importance-aware scheduling with label information  is compared with the unlabeled scheduling scheme and the two benchmarks.  By exploiting additional label information,  the model attains faster convergence rate than the unlabeled scheme.  However, if the transmission budget is large, the learning accuracies of two schemes will be converged to a comparable level.  Comparing with the benchmarks, the importance-aware scheduling with label information achieves remarkable improvement in terms of learning accuracy, corresponding to  5$\%$ for channel-aware scheduling and 8 $\%$ for data-aware scheduling.   }

 \vspace{-25pt}

\subsection{Learning Performance for CNN}
\begin{figure}[t]
\begin{center}
{\includegraphics[width=8cm]{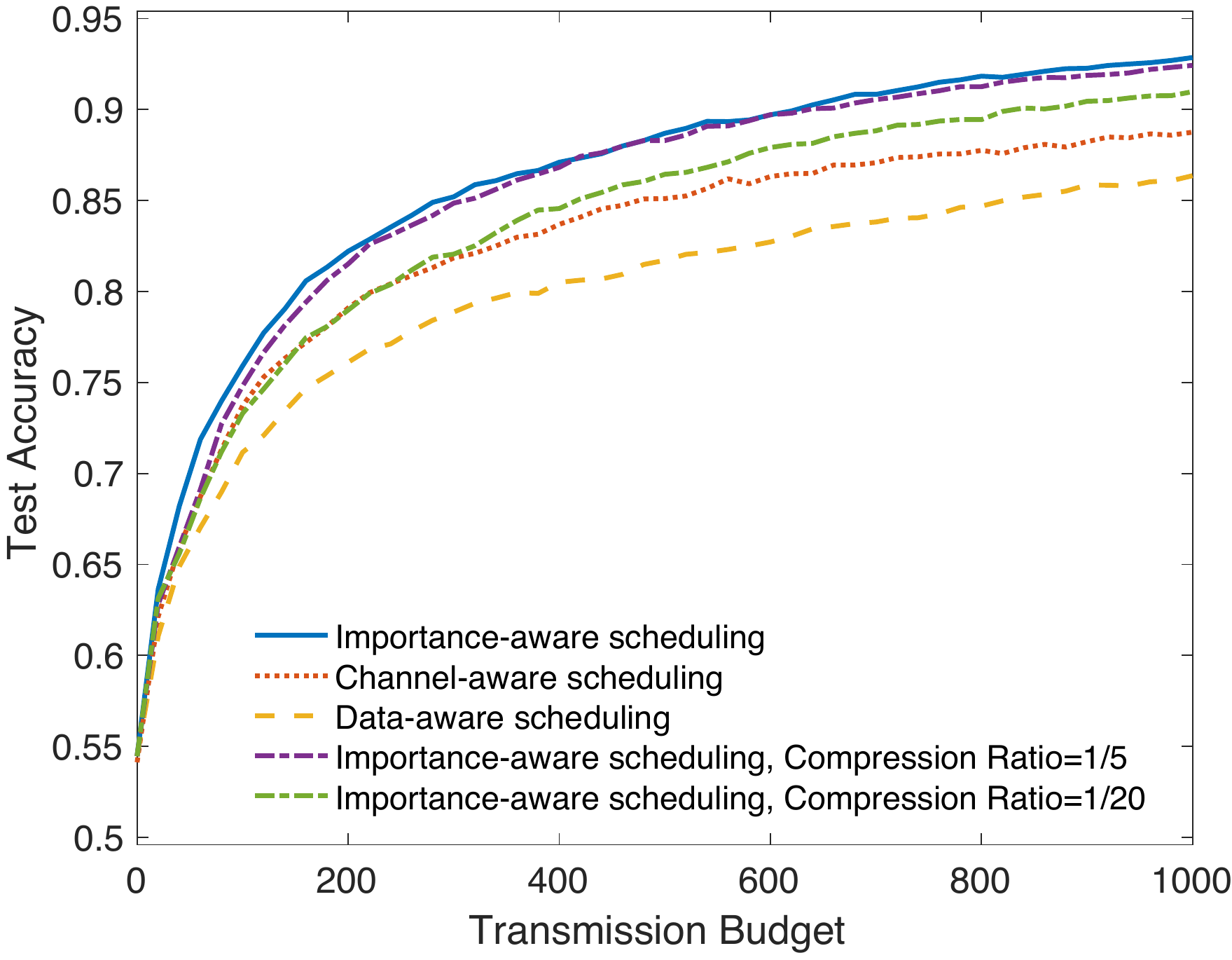}}\\
\caption{Learning performance evaluation for CNN classifier.}
\label{Fig: CompressionCNN}
\end{center}
 \vspace{-20pt} 
\end{figure}
Our heuristic design for CNN is tested in the scenario of multi-class classification and the results are provided in Fig.~\ref{Fig: CompressionCNN}.  For the uncompressed evaluation model, the test accuracy consistently outperforms the two baseline schemes.  Although the heuristic design can not guarantee an optimal tradeoff between data importance and data (channel) reliability, the performance gain is prominent.  That confirms the benefit of exploiting both channel-and-data diversity in CNN.  On the other hand, each training sample in CNN contributes to define the multiple decision boundaries, that is different from SVM where a single hyperplane determined by few support vectors.  In another word, CNN is less robust with the existing of channel noise, and requires more training samples to construct multiple boundaries.  Thereby it has a potential to achieve more remarkable performance gain if more users are involved in the system, since the achievable multiuser diversity for SVM may not be enough for the training of CNN classifier.

In general, a large number of CNN parameters are redundant that enables a highly compressed model for importance evaluation.  This can be supported by the curve of  $C_r=1/5$, which achieves almost same performance as that of by using the uncompressed one.  Furthermore, the performance loss will be less than $50\%$  if the compression ratio is $1/20$. In contrast, the SVM classifier suffers a prominent performance loss for  $C_r=1/10$.  That verifies the redundancy of CNN which is capable to utilize a compressed evaluation model to reduce computing complexity.

\section{Concluding Remarks}\label{sec: concluding remarks}

In this paper, we have proposed the novel scheduling scheme, namely importance-aware scheduling, for wireless data acquisition in edge learning systems.  The scheme intelligently makes a joint channel-and-data selection for training data uploading so as to accelerate learning speed. Comprehensive experiments using real datasets substantiate the performance gain by exploiting two-fold multi-user diversity, namely multiuser data-and-channel diversity.  
  
At a higher level, the work contributes the new principle of exploiting data importance to improve the efficiency of multiuser data acquisition for distributed edge learning. This work can be generalized to a more sophisticated batch mode training, where multiple devices can be scheduled in a broadband system with OFDMA or a MIMO system with SDMA. The scheduling algorithms can be further designed to ensure global data diversity, where the feedback is required to account for both multiuser diversity and diversity with respect to global dataset.  Besides raw data acquisition, another interesting direction is the acquisition of  learning relevant information in a federated learning framework, e.g., gradient updates and model updates, and the design therein involves changing the data-importance metrics to gradient divergence and model variance.

\bibliographystyle{ieeetr}

\end{document}